\documentclass[pra,letterpaper,aps,10pt,superscriptaddress,twocolumn,floatfix,showpacs]{revtex4-2}
\usepackage{graphicx}
\usepackage{amsmath}
\usepackage{amsfonts}
\usepackage{empheq}
\usepackage{float}
\usepackage{amssymb}
\usepackage{epsfig}
\usepackage{amsmath}
\usepackage{epstopdf}
\usepackage{placeins}
\DeclareGraphicsExtensions{.pdf,.eps,.png,.jpg,.mps}
\usepackage[pdftex]{color}
\usepackage{amsmath,graphicx,amssymb,braket,subfigure,upgreek}
\usepackage[dvipsnames]{xcolor}
\usepackage[colorlinks, linkcolor=blue, citecolor=blue, urlcolor=blue, breaklinks=true]{hyperref}
\usepackage{microtype}
\usepackage{bbm}
\usepackage{color}
\usepackage{dsfont}

\begin{document}

\title{Spatial delocalization of cold oscillators in hollow-core fibers}
% \title{Adaptive optomechanical control of a fringe-trapped nanoparticle in a hollow-core fiber}
\author{Soumya Chakraborty}
\email{soumya.chakraborty@mpl.mpg.de}
\affiliation{Department of Physics, Friedrich-Alexander-Universit\"{a}t Erlangen-N\"urnberg, Staudtstra{\ss}e 7,
D-91058 Erlangen, Germany}
\affiliation{Max Planck Institute for the Science of Light, Staudtstra{\ss}e 2,
D-91058 Erlangen, Germany}
\author{Pardeep Kumar}
\email{pardeep.kumar@mpl.mpg.de}
\affiliation{Max Planck Institute for the Science of Light, Staudtstra{\ss}e 2,
D-91058 Erlangen, Germany}
\author{Gordon K. L. Wong}
\email{gordon.wong@mpl.mpg.de}
\affiliation{Max Planck Institute for the Science of Light, Staudtstra{\ss}e 2,
D-91058 Erlangen, Germany}
\author{Claudiu Genes}
\email{claudiu.genes@physik.tu-darmstadt.de}
\affiliation{TU Darmstadt, Institute for Applied Physics, Hochschulstra{\ss}e 4A, D-64289 Darmstadt, Germany}
% \affiliation{Max Planck Institute for the Science of Light, Staudtstra{\ss}e 2,
% D-91058 Erlangen, Germany}
%\affiliation{Department of Physics, Friedrich-Alexander-Universit\"{a}t Erlangen-N\"urnberg, Staudtstra{\ss}e 7,
%D-91058 Erlangen, Germany}
\author{Nicolas Y. Joly}
\email{nicolas.joly@fau.de}
\affiliation{Department of Physics, Friedrich-Alexander-Universit\"{a}t Erlangen-N\"urnberg, Staudtstra{\ss}e 7,
	D-91058 Erlangen, Germany}
\affiliation{Max Planck Institute for the Science of Light, Staudtstra{\ss}e 2,
D-91058 Erlangen, Germany}
\date{\today}

\begin{abstract}
Trapped particles in hollow-core fibers enable long-range sensing, though advanced control of their in-fiber motion, such as multimodal cooling and squeezing, remains challenging. Here, we present optical interference-based adaptive techniques for controlling the motion of a fringe-trapped silica nanoparticle inside a fiber. After feedback-cooling its axial and radial motion, we induce axial delocalization (position anti-squeezing) via two complementary approaches. First, non-adiabatic fringe suppression expands the position variance by 11.83 (±0.7) dB to that of the initial cold-state while retaining Gaussian statistics. Second, multi-pass particle positioning at dark fringes increases the delocalization to 13.23 (±0.5) dB relative to the cold-state’s variance via dark inverted optical potentials, in agreement with our Wiener stochastic model. Stronger delocalization produces non-Gaussian states of motion. Unlike prior inverted-trap implementations, our method requires neither auxiliary optical traps nor charged particles in Paul traps. Our results demonstrate fringe-trapped particles in hollow-core fibers as a versatile platform for long-range sensing and macroscopic quantum physics.
\end{abstract}


\maketitle

\section{Introduction}
Optomechanics~\cite{GonzalezBallestero2021Levitodynamics,Florian_review} with optically trapped nanoparticles is opening new opportunities in metrology and tests of quantum physics, including gravity-induced wavefunction collapse. Recent advances include ground-state cooling~\cite{Genes, delicground,Novotnyground, Magrini2021}, quantum-limited force sensing~\cite{force_zero_point, kamba_force}, and preparation of atypical motional states~\cite{vorvosh,northrup_sq,ruffato,Song2026,Zhang2026TwoModeThermomechanical}.

Building on these foundations, highly delocalized motional states, characterized by anti-squeezing along the position quadrature, provide a pathway toward surpassing standard quantum limits in force sensing~\cite{kamba_force} and preparing nanoparticles in macroscopic quantum superpositions~\cite{romero_pra_2011}. In optical tweezers~\cite{Ashkin69}, motional delocalization is typically achieved through non-adiabatic modulation of the trapping power, which changes the mechanical frequency~\cite{vorvosh,Zhang2026TwoModeThermomechanical,gieseler} and has yielded up to 7.2 dB of quantum squeezing~\cite{delocalisation,Kamba2025QuantumSqueezing}. Dark inverted traps offer an alternative path to squeezing with near-zero quantum backaction and rapid state expansion, but existing implementations rely on auxiliary traps~\cite{kremer2026fast,Duchan2025} and charged particles in Paul traps which are sensitive to stray electric fields~\cite{romain-dark}.

These limitations motivate alternative architectures, such as standing-wave fringe traps formed by interfering counterpropagating fiber modes in hollow-core photonic crystal fibers (HC-PCFs). These diffraction-free traps~\cite{Bykov,Chakraborty:26,grass} enable long-range particle transport~\cite{linder}, while the relative phase and polarization of the counterpropagating modes independently control the particle position and axial stiffness. We recently demonstrated axial feedback cooling through the relative phase control~\cite{Kumar,cg5v-lpjy}; however, advanced control, including multidimensional cooling and squeezing, remains unexplored.

Here, we demonstrate adaptive control of the dissipative and conservative optomechanical dynamics of a 234-nm silica nanoparticle trapped in an HC-PCF under high vacuum. We combine phase-adaptive cold damping of the axial mode with parametric feedback of the radial mode~\cite{Kumar,cg5v-lpjy,ghosh2023phase,gieseler}, reducing their center-of-mass (CoM) temperatures near 1~K at 10~$\mu$bar pressure. After cooling the particle's motion, we demonstrate two complementary protocols for axial thermomechanical delocalization. A sudden polarization change produces a nonadiabatic axial frequency jump \textit{without} changing the trapping power, increasing the position variance of the state by 11.83~($\pm0.7$)~dB relative to the variance of a cold (7.8 K) thermal state while preserving Gaussianity. This delocalized state subsequently relaxes toward the initial thermal state under thermal noise and damping, as accurately captured by our stochastic Wiener evolution. Independently, rapid phase control transports the particle to and fro between bright and dark optical fringes, where the dark inverted potential produces hyperbolic motional evolution and exponential phase-space delocalization. 
Repeated phase-synchronized excursions between dark and bright fringes increases the position variance of the thermal state by 13.23~($\pm0.5$) dB relative to the initial state's variance, \textit{without} requiring auxiliary traps~\cite{Duchan2025} or charged particles~\cite{romain-dark}. Beyond the linear regime, further increases in dark fringe excursion delocalize the particle enough to probe the nonlinearities of the optical fringe potential, giving access to non-Gaussian motional dynamics~\cite{ruffato}.

The demonstrated delocalized thermal states open a unique route for precision force and displacement sensing~\cite{kamba_force} with trapped nanoparticles. In our system, the achieved $\sim 13~\mathrm{dB}$ squeezing of velocity variance corresponds to a 4.5-times enhancement in force sensitivity when compared to the original cold thermal state. The dark fringe particle placement also enables rapid delocalization of the particle's CoM wavepacket and can be useful for matter-wave interferometry~\cite{romero_pra_2011} with motional cat states of increasingly massive objects. This connection between adaptive control, precision sensing, and quantum state engineering establishes HC-PCF-based fringe trapping as a versatile platform for fiber-integrated metrology and optomechanics.

\section{Cooling action} 
\begin{figure}[ht!]
 \includegraphics[width=\columnwidth]{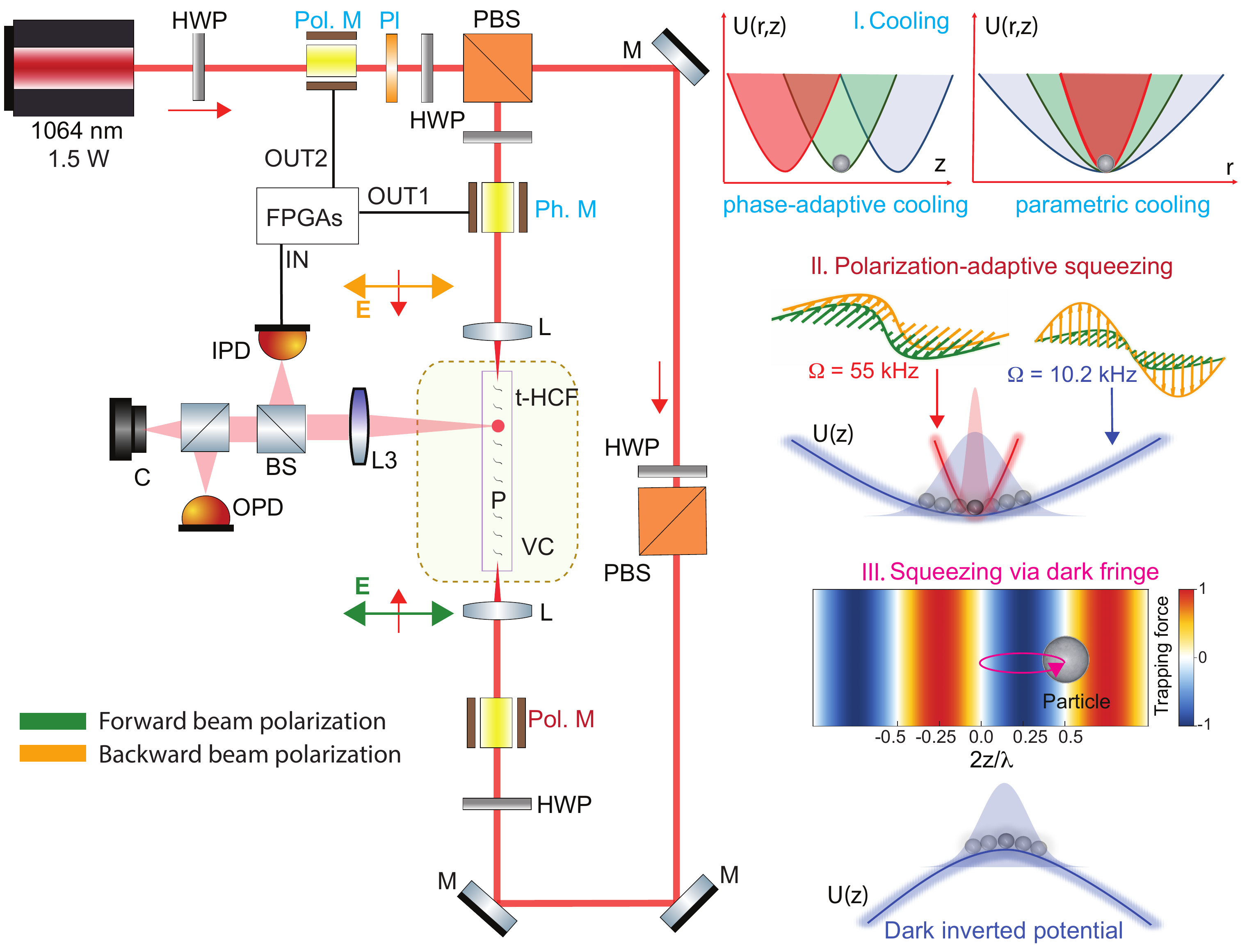}
\centering % Centers the figure in the column
\caption{\textbf{Schematic of the HC-PCF dual-beam optical tweezer for multimode cooling and axial delocalization;} P: levitated particle, HWP: half-wave plate, PBS: polarizing beam splitter, Pl: Polarizer, BS: beam splitter, C: camera,  M: mirror, VC: vacuum chamber, t-HCF: twisted hollow-core fiber, L: 75 mm lens, L3: 0.26 NA objective, OPD: out-of-loop photodiode, IPD: in-loop photodiode, Pol. M: Polarization modulator, Ph. M: phase modulator, FPGAs: Moku:Lab and Red Pitaya STEMlab-125. The forward (left $\rightarrow$ right direction) and backward (right $\rightarrow$ left direction) beams' polarizations (electric field orientations) are shown in green and yellow colors, respectively. Active feedback cools the axial motion by modulating the standing-wave position $($position of the axial potential $U(z))$ via the relative phase of the counterpropagating beams, and the radial motion by modulating the radial trap stiffness $($slope of the radial potential $U(r))$. Mechanical anti-squeezing is realized either by transiently suppressing the standing wave using nearly-orthogonally polarized trapping beams which reduces the axial stiffness or by momentarily placing the particle at a dark fringe, therefore creating an inverted axial optical potential.}
\label{fig2}
\end{figure}

A schematic of the experimental setup is shown in Fig.~\ref{fig2}. A 1064~nm ($\lambda$) Nd:YAG laser with 1.5~W total trapping power is split into two equally intense, co-polarized beams and coupled into the fundamental modes of a twisted hollow-core fiber (t-HCF)~\cite{cg5v-lpjy} with a 44~$\mu$m core diameter. The fiber is mounted inside a vacuum chamber to control the gas damping of the trapped particle's motion. Scattered light from the particle, together with the cladding stray light, is collected by a 0.26 NA objective and detected using a camera and two fast photodiodes. The photodiode signals are obtained through interference between particle-scattered and cladding light~\cite{cg5v-lpjy}. The in-loop photodiode (IPD) provides the signals for field-programmable gate-array (FPGA)-based feedback control, while the out-of-loop photodiode (OPD) is used for independent motion detection. The particle is loaded at atmospheric pressure by nebulizing a highly diluted silica suspension. At around 5~mbar, the axial mode is observed at $\Omega=2\pi\times55\rm ~kHz$. The radial mode is at $\Omega_r=2\pi\times1.3~\rm kHz$. The motional signals are calibrated using equipartition theorem~\cite{Li2013}. 

\begin{figure}[htbp]
    \centering
    \includegraphics[width=\columnwidth]{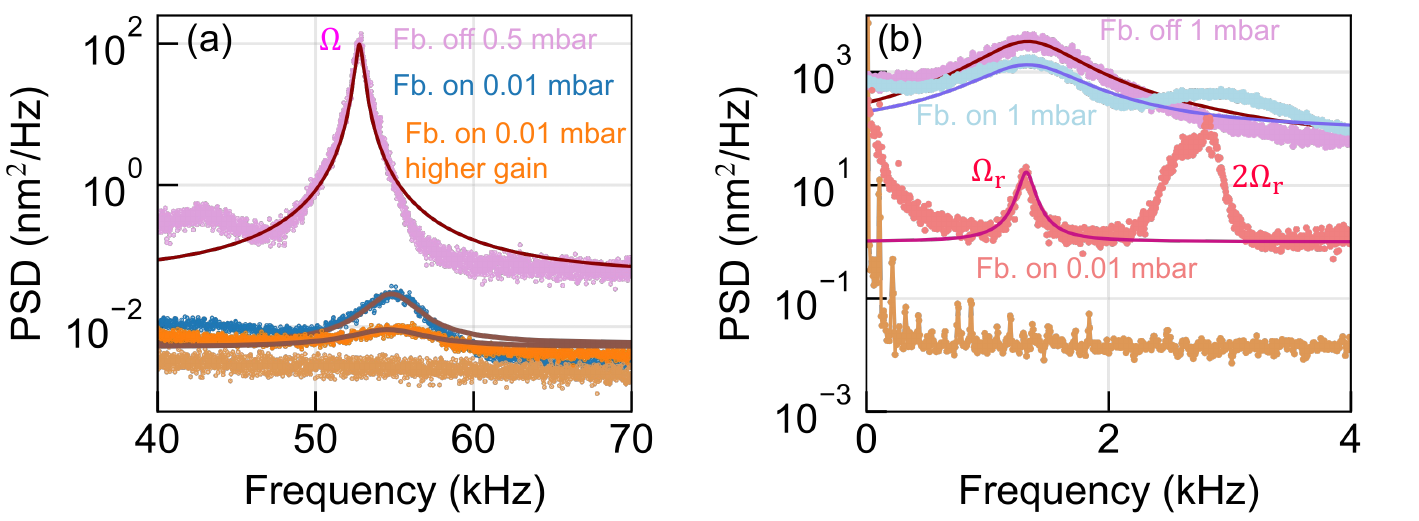}
    \caption{\textbf{Simultaneous cooling of two eigenmodes;} (a) Axial PSD shown at 0.5 mbar without feedback cooling (plum colored). The
   blue and orange PSD at 0.01 mbar are recorded with feedback cooling for two different feedback cooling rates of 4 kHz and 7.2 kHz, respectively.  (b) Parametric feedback cooling of the radial eigenmode of a 234 nm silica particle. The plum-colored spectra are recorded in the absence of parametric cooling at 1 mbar of air pressure. The steel-blue and coral-red-colored spectra are recorded at 1 mbar and 0.01 mbar air pressure in the presence of feedback cooling featuring second harmonic feedback signals at $2\Omega_r$. All the spectra are fitted with Lorentzian curves. }
    %The temperature of the axial and radial eigenmodes are 0.7 K and 2.9 K respectively.}
    \label{radial_cooling_spectra}
\end{figure}

The IPD signal, which contains both axial and radial eigenmodes, is digitally filtered in real time at the respective eigenfrequencies with sufficient bandwidth. For axial cooling, the Red Pitaya FPGA uses its IQ module to generate a feedback signal at the axial frequency $\Omega$, with a phase shift of $\pi/2$ relative to the measured motion. This signal drives a phase modulator that dynamically adjusts the relative phase of the counterpropagating trapping beams, producing a viscous damping force on the axial CoM motion. For radial cooling [Fig.~\ref{radial_cooling_spectra}(a)], the Moku:Lab FPGA generates a phase-synchronized modulation signal at \(2\Omega_r\) with appropriate damping using its phasemeter module. This signal drives an amplitude modulator to parametrically control the trap stiffness, increasing the confinement as the particle moves away from the trap center and reducing it during the return motion, resulting in net damping of the radial eigenmode. As shown in Fig.~\ref{radial_cooling_spectra}(b), the radial motional peak appears at $\Omega_r$ and the power modulation appears at $\sim2\Omega_r$. The power-modulation peak is broader than the radial peak because the feedback signal decays twice as fast as the radial motion. This is because the feedback modulation follows the CoM energy, which oscillates and decays twice as fast as the position oscillation~\cite{gieseler}. At a given pressure, axial cold-damping cooling is approximately five times more efficient than radial cooling as measured through feedback-induced broadening of the corresponding spectral linewidths. The axial and radial temperatures are reduced to \(0.7\)~K and \(2.9\)~K, respectively, at \(10~\mu\mathrm{bar}\) pressure (Fig.~\ref{radial_cooling_spectra}). The particle remains stably trapped down to \(10~\mu\mathrm{bar}\) which is the current limit of our vacuum system. The temperatures are measured by comparing the area under the feedback-cooled spectra to the spectra in the absence of cooling. The noise floor obtained for axial trapping frequency is 49 pm/$\sqrt{\mathrm{Hz}}$. The slight change in trapping frequencies are most likely caused by trapping laser power fluctuations and power modulation over the duration of experiment~\cite{gieseler}.

\section{Delocalization}
From the cold thermal state we then engineer, using different protocols, an axially delocalized harmonic oscillator. The delocalization experiments are performed at a slightly higher pressure and an axial CoM temperature of 7.8 K to increase the squeezed-state decay time, which scales inversely with pressure, thereby reducing the measurement time.

\subsection{Polarization-Adaptive Delocalization}
 In the first protocol, we prepare a delocalized thermal state by impulsively reducing the axial optical confinement through rotating the relative polarization angle ($\theta_p$) between the trapping beams from $0^\circ$ to $\sim 88.3^\circ$. The axial frequency depends on the relative polarization angle as $\Omega=\Omega_0\sqrt{\cos\theta_p}$ [see Appendix \ref{appendixB}]. At $\theta_p=88.3^\circ$, the axial and radial frequencies decrease to $10.2$ kHz and $0.92$ kHz, respectively~\cite{Chakraborty:26}. The shift in the radial eigenfrequency indicates that polarization rotation also modifies the radial trap, consistent with an approximately $1/\sqrt{2}$ reduction in the total on-axis electric-field intensity [see Appendix \ref{appendixB}]. The time scale of the switch between collinear to nearly-orthogonal polarization states of the trapping beams is determined by the electro-optic modulator bandwidth, which is more than 100 MHz and therefore much larger than the trap frequencies (kHz). 

For axial delocalization, the feedback is then switched off after cooling the particle, and the standing-wave trap is transiently suppressed, reducing the axial frequency to $\Omega_1=2\pi\times10.2$~kHz. The particle evolves in the shallow potential for one-quarter of the corresponding oscillation period, producing motional delocalization through the sudden frequency change. The standing-wave trap is subsequently restored, and feedback cooling is reactivated to reinitialize the particle. This sequence is repeated 3000 times, while the particle position, extracted from the OPD signal and real-time filtered around the axial frequency using the Red Pitaya, is continuously recorded. The velocity is obtained by retrodicting the measured position~\cite{rossi_retro,northrup_sq, romain-dark}, yielding the motional phase-space trajectory. Further details regarding the data processing are provided in Appendix \ref{appendixA}.

\begin{figure}[htbp]

    \centering
\includegraphics[width=\linewidth]{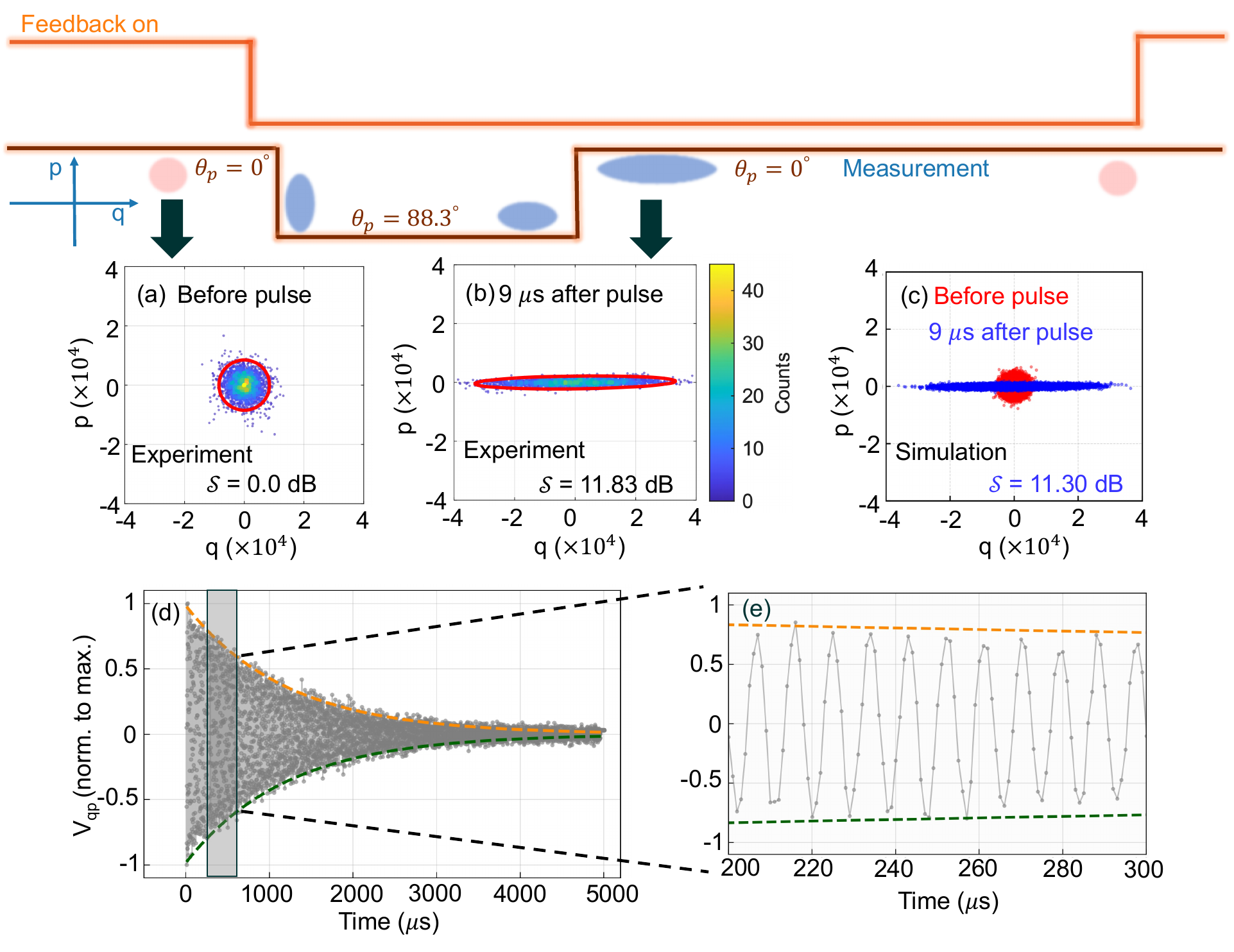}
    \caption{\textbf{Evolution of delocalized thermal states;} Experimental (a-b) and simulated (c) phase-space distributions normalized to the zero-point motion (zpm), showing the anti-squeezed state under thermal noise. The position and momentum zpms are $q_{\rm zpm}=4.43 \rm ~pm$ and $p_{\rm zpm}=22.5 \rm ~yN.s$. The red circle (ellipse) on top of (a) and (b) are the analytical distributions using Eq.~\eqref{eq_Prob_dist}.  The experimental data and the numerical data are taken over 3000 and 20000 realizations, respectively. (d-e) Thermal decoherence of the position–momentum correlation ($V_{qp}$), shown at times starting 9~$\mu\mathrm{s}$ after application of the polarization-rotating pulse. The time $0~\mu \rm s$ refers to the end of the pulse. The red and blue envelop are the fit to the gas damping~\cite{Li2013}. (e) is the magnified view of (d).}
    \label{squeezing1}
\end{figure}

The feedback-cooled particle initially has a circular phase-space distribution as shown in Fig.~\ref{squeezing1}(a), where potential and kinetic energies are balanced. A sudden change of the trap stiffness via relative polarization rotation rapidly changes the potential energy without changing the particle’s instantaneous motion. The subsequent evolution converts this imbalance into correlated position and momentum quadratures, stretching the thermal distribution into an ellipse. This redistribution of fluctuations between the two quadratures produces an anti-squeezed motional state.
We observe up to 11.83~($\pm 0.7$)~dB of delocalization of the trapped particle, as shown in Fig.~\ref{squeezing1}(b), inferred from the phase space dynamics $9 ~\mu\rm s$ after the two frequency jumps ($\Omega_0\rightarrow\Omega_1\rightarrow\Omega_0$). The uncertainty is computed by fitting the position distributions with a Gaussian function. At this point, we observed a strong correlation between the position and momentum of the trapped particle $V_{qp}$ which decays as a function of time due to collision from the background gas molecules as shown in Fig.~\ref{squeezing1}(d-e).

We describe the measured dynamics  by the covariance matrix $V(t)=\langle\mathbf{u}(t)\mathbf{u}^{T}(t)\rangle$, where $\mathbf{u}=(q,p)^T$ denotes the vector of dimensionless position and momentum quadratures, and $\langle...\rangle$ denotes the ensemble average over independent realizations of the stochastic dynamics. The system evolves under deterministic dynamics and stochastic thermal fluctuations [see Appendix \ref{appendixC}]. Starting from thermal equilibrium at trap frequency $\Omega_0$, we apply a square polarization pulse that rapidly changes the
trap frequency from $\Omega_0$ to $\Omega_1$ at $t_1$ and restores it to $\Omega_0$ at $t_2$. Since both polarization rotations are fast compared with the mechanical period, they act as instantaneous symplectic
transformations. The first quench maps the covariance as $V(t_1^+)=S(r)V(t_1^-)S^T(r)$, with $S(r)=\operatorname{diag}(e^r,e^{-r})$ and $r=\ln(\sqrt{\Omega_1/\Omega_0})$, where and throughout the analysis the superscripts $\mp$ denote instants immediately before and after the quench, respectively. It thereby transforms the initially
isotropic thermal distribution into an elliptical phase-space distribution.

During the pulse, the covariance ellipse rotates at $\Omega_1$, while
damping and thermal diffusion progressively reduce its anisotropy,
\begin{equation}
V(t)=e^{-\gamma\tau_{1}}\mathcal{R}(\Omega_1\tau_{1})V(t_1^+)
\mathcal{R}^{T}(\Omega_1\tau_{1})+\mathcal{Q}_1(\tau_{1})\;,
\label{eq_Cov_final}
\end{equation}
where $\tau_{1}=t-t_1$, $\mathcal{R}$ describes harmonic phase-space rotation and $\mathcal{Q}_1$ is the thermal covariance. To enhance anisotropy upon the return quench, we set the pulse duration to $T_1/4$ ($T_1=2\pi/\Omega_1$). This quarter-period evolution eliminates the position–momentum correlation, leaving $V(t_2^-)$ diagonal with unequal variances. The inverse transformation $V(t_2^+)=S(-r)V(t_2^-)S^T(-r)$ then amplifies this asymmetry. Subsequent evolution at $\Omega_0$ follows Eq.~\eqref{eq_Cov_final} with $\Omega_1\to\Omega_0$ and $\mathcal{Q}_1\to\mathcal{Q}_0$. Following this sequence, the transient dynamics as shown in [Fig.~\ref{squeezing1}(d)] feature a coherently rotating ellipse whose anisotropy decays to equilibrium via damping and diffusion.

Diagonalizing the post-pulse covariance matrix yields the variances of the squeezed and anti-squeezed quadratures, $\Lambda_{\min}(t)$ and $\Lambda_{\max}(t)$. We quantify the transient position anti-squeezing (delocalization) in decibels as \cite{wollman_Squeeze_def_Science_2015},
\begin{equation}
    \mathcal{S}(t) = -10\log_{10}\left[\frac{\Lambda_{\mathrm{min}}(t)}{n_\mathrm{th}}\right]\;,
    \label{eq_Squeezing}
\end{equation}
where $n_{\rm th}$ is the initial thermal variance. Including damping and thermal noise, the minimum variance immediately following the pulse is 
\begin{align}
    \Lambda_{\min} &=e^{-\gamma T_{1}/4}n_{\mathrm{th}}e^{-4|r|}+n_{\mathrm{th}}^{\Omega_{1}}(1-e^{-\gamma T_{1}/4})e^{-2|r|}\;,\label{eq_lambda_min}
\end{align}
where $n_{\mathrm{th}}^{\Omega_{1}}$ is the phonon occupation at $\Omega_{1}$. For our single pulse protocol (frequency jumps from 55 to 10.2 kHz, back to 55 kHz), Eq.~\eqref{eq_Squeezing} predicts $11.96$~dB of delocalization immediately after the pulse as shown in Fig.~\ref{squeezing1}(c). At longer times, damping and thermal diffusion erase this anisotropy, restoring equilibrium and returning squeezing to zero. The numerical simulation and analytics involving the thermal noise and damping shows excellent agreement with the experimental results. 

Because the particle remains confined to a single fringe, the potential is well approximated as harmonic and the dynamics remain linear. The phase-space distribution remains Gaussian and is fully characterized by the covariance matrix, 
\begin{equation}
    P(q,p,t)=
    \frac{1}{2\pi\sqrt{\det V(t)}}
    \exp\left[
        -\frac{1}{2}
        \mathbf{u}^{T}V^{-1}(t)\mathbf{u}
    \right]\;,\label{eq_Prob_dist}
\end{equation}
providing a direct connection between the analytical covariance dynamics, which are shown in the red-colored ellipses in Fig.~\ref{squeezing1}, and  experimentally observed distributions.

\subsection{Dark Fringe-Adaptive Delocalization}
An alternative to polarization-based delocalization is to place the particle at a dark fringe of the standing-wave trap, where the local intensity and hence recoil scattering are strongly suppressed, while the inverted optical potential enables accelerated squeezing. To generate anti-squeezing in the dark trap, we transiently shift the standing-wave pattern by $\lambda/4=266$ nm, transferring the particle from a bright to a dark fringe. The particle evolves in the dark potential for $1.8~\mu$s before the standing wave is shifted back, returning the particle to the bright fringe and preventing particle loss. After evolving for half a motional period, this dark-fringe excursion is repeated to accumulate squeezing. Each individual excursion from $\rm bright \rightarrow dark \rightarrow bright$ fringes is referred to as `dark pulse'. The motional state is measured when the particle is in the bright fringe after each excursion, followed by feedback cooling to reinitialize the system. The full sequence is repeated 3000 times.

Building on the frequency-quench description, we now describe a complementary anti-squeezing route with a fixed axial frequency  $\Omega$, where phase-adaptive control rapidly transfers the particle from the bright to the dark fringe for duration $\tau_D$. This fast transfer leaves the covariance continuous across the transition, $V(t_1^-)=V(t_1^+)$. At the dark fringe, the inverted potential amplifies displacements, so that the dynamics are described by the hyperbolic propagator
\begin{equation}
    \mathcal{O}_{D}(\Omega\tau_D)
    =
    \begin{bmatrix}
        \cosh(\Omega\tau_D) & \sinh(\Omega\tau_D)\\
        \sinh(\Omega\tau_D) & \cosh(\Omega\tau_D)
    \end{bmatrix}\;,
    \label{eq_dark_propagator}
\end{equation}
whose eigenvalues $e^{\pm\Omega\tau_D}$ define expanding and contracting directions. Including damping and thermal fluctuations, the covariance in the dark-fringe evolves as
\begin{equation}
V(t)
=
e^{-\gamma\tau_D}
\mathcal{O}_{D}(\Omega\tau_D)
V(t_1^+)
\mathcal{O}_{D}^{T}(\Omega\tau_D)
+
\mathcal{Q}_{D}(\tau_D)\;,
\label{eq_dark_covariance}
\end{equation}
with $\mathcal{Q}_{D}$ capturing the thermal noise [see Appendix \ref{appendixD}]. The dark interval is kept finite---sufficient for substantial delocalization, yet short enough to prevent too large displacements. Upon return to the bright fringe, the covariance remains unchanged, and the subsequent bright interval of duration $\tau_B$ evolves under  Eq.~\eqref{eq_Cov_final}.

For a train of $N_p$ dark-fringe pulses separated by bright intervals $\tau_B$, the covariance $V_j$ after the $j$-th pulse ($j=1,\dots,N_p-1$) obeys the recursive relation
\begin{equation}
    V_{j+1}
    =
    e^{-\gamma(\tau_D+\tau_B)}\mathcal{M} V_j \mathcal{M}^{T}
    +
    \mathcal{Q}_{\mathrm{cyc}}\;,
    \label{eq_dark_coavariance_pulses}
\end{equation}
where $\mathcal{Q}_{\mathrm{cyc}}$ collects the thermal covariance accumulated per cycle and $\mathcal{M} = \mathcal{R}(\Omega\tau_B)\mathcal{O}_{D}(\Omega\tau_D)$ represents a single dark-bright cycle. This map captures the phase-sensitive accumulation: $\tau_D$ controls squeezing strength, while $\tau_B$ sets the orientation of the anti-squeezed ellipse for constructive addition.  From this recursion, anti-squeezing is quantified by Eq.~\eqref{eq_Squeezing}, and the linear-Gaussian form of the distribution is preserved throughout the sequence [Eq. \eqref{eq_Prob_dist}]. 

\begin{figure}[htbp]
    \centering
\includegraphics[width=\linewidth]{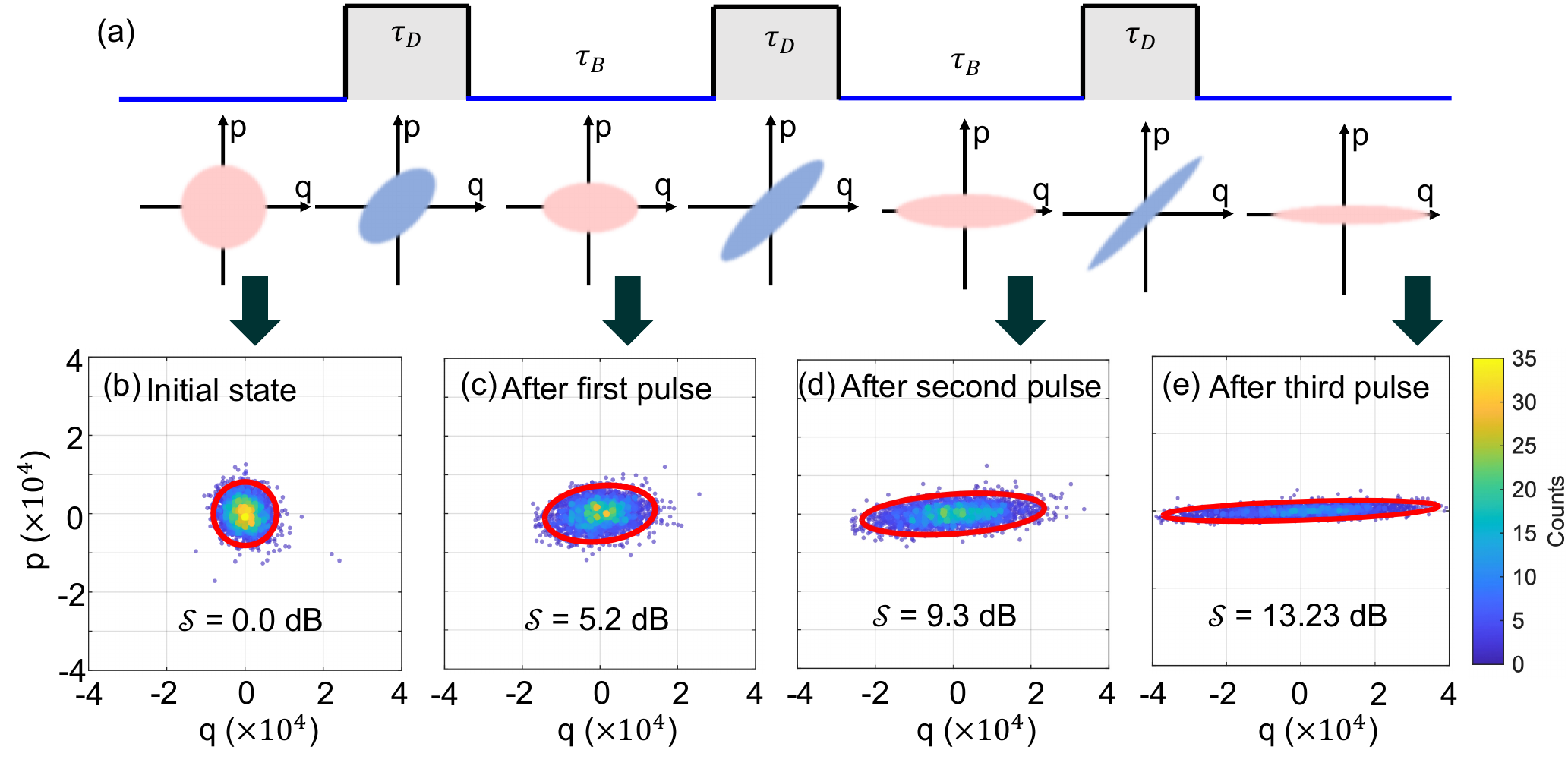}
\caption{\textbf{Demonstration of nanoparticle motional delocalization using a transient inverted optical potential;} (a--e) The standing-wave trap is shifted by to the dark fringe for three times in a row and measurement is done once the particle is transferred back to the bright region, producing an increase of the position variance by 13.23~($\pm0.5$) dB relative to the initial thermal variance. The bright and dark fringe particle placement durations are $\tau_D=1.8 ~\mu \rm s$ and $\tau_B=9 ~\mu \rm s$, respectively. The experimental  distributions are normalized to the zero-point  motion (zpm) and fitted with Eq.~\eqref{eq_Prob_dist}.}
    \label{squeezing_dark}
\end{figure}

During the $1.8~\mu\mathrm{s}$ interaction time, this inverted potential induces exponential stretching and compression of the motional phase-space distribution [see Fig.~\ref{squeezing_dark}(a)]. To prevent the particle from reaching an adjacent bright fringe as the position uncertainty grows, which would halt the intended exponential delocalization without a clear signature, we periodically return the particle to the bright fringe before transferring it back to the dark fringe. The period is chosen to be half the oscillation period, such that the delocalization generated during each dark-fringe trapping event accumulates over subsequent events, as shown in Fig.~\ref{squeezing_dark}(c, d \& e). This repositioning the particle at the dark fringes across multiple passes transiently increases the position variance up to $13.23~(\pm0.5)~\rm dB$ relative to the initial thermal variance.

\section{Non-Gaussian State Preparation}
\begin{figure}[htbp]
    \centering
\includegraphics[width=0.8\linewidth]{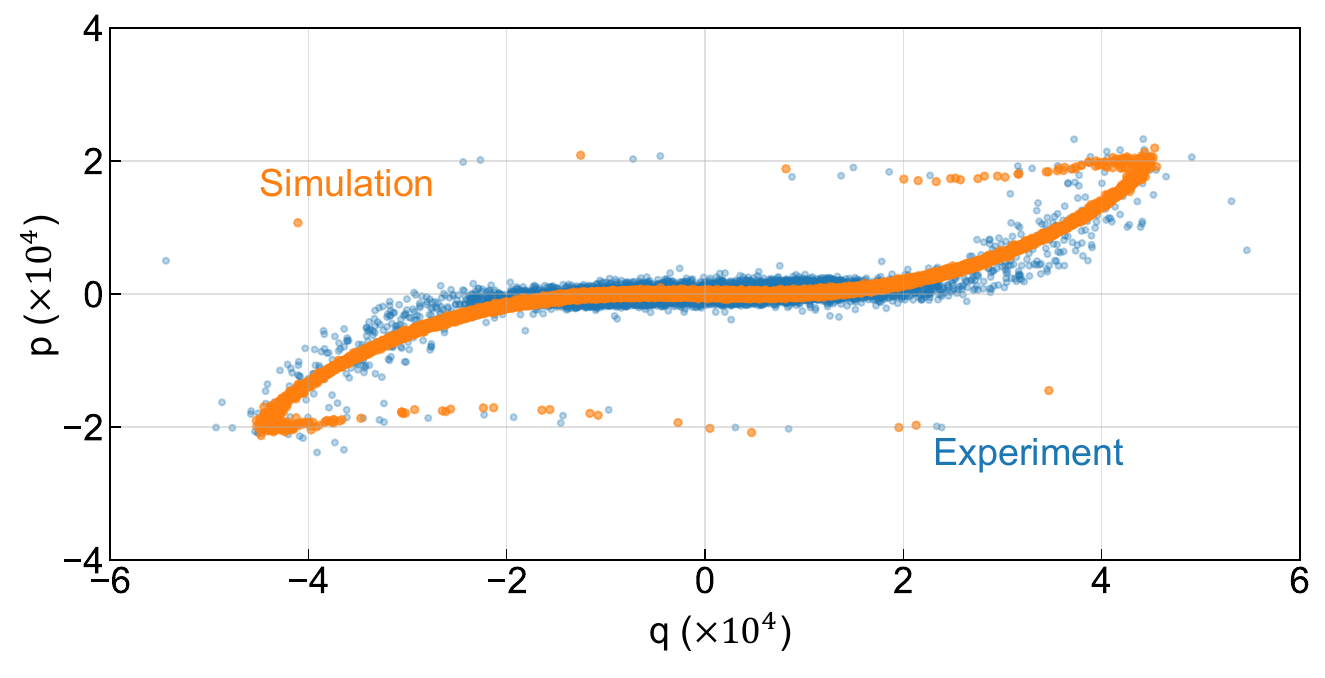}
\caption{\textbf{Non-Gaussian motional state preparation;} Numerical and experimental phase-space distributions of a non-Gaussian state generated through massive delocalization in the nonlinear optical trap using four consecutive dark-fringe placement events. The blue points are experimental data taken over 6000 realizations while the orange points are numerical solution of nonlinear motional dynamics taken over 20000 realizations.}
    \label{nonlinear}
\end{figure}

With increasing position delocalization, the cold particle reaches the fringe-trap edges, where quartic and higher-order even terms introduce significant anharmonicity into the optical potential, modifying the restoring force and particle dynamics [see Appendix \ref{appendixE}]. To capture this nonlinearity, we retain the full spatial dependence of the standing-wave intensity~\cite{Kumar} rather than expanding about the trap minimum, giving a sinusoidal gradient force. Fig.~\ref{nonlinear} compares experiment and theory: a fourth dark pulse measures the motional state at $T/8\simeq2 ~\mu \rm s$ after the particle returns to the bright fringe (blue data), while numerical solution of the nonlinear stochastic equations [Eqs.~\eqref{eq_nonlinear_q}-\eqref{eq_nonlinear_p} in Appendix \ref{appendixE}] of motion simulates the non-Gaussian state (orange data). Fitting the simulation to the experimental data gives the nonlinearity parameter $\eta = 2.69\times10^{-5}$, consistent with the theoretical expression $\eta = 2\pi q_{\mathrm{zpm}}/\lambda$, where $q_{\rm zpm}=\sqrt{\hbar/m\Omega}=4.43~\rm pm$ is the zero-point position fluctuations with $\hbar$ being the reduced Planck constant.

As the highly anti-squeezed thermal state evolves in the bright anharmonic optical potential, its phase-space distribution develops extended, curved, and fanning tails. While low-amplitude components of the distribution remain predominantly confined to the approximately harmonic region of the potential, higher-amplitude components experience an amplitude-dependent oscillation frequency. Consequently, different parts of the distribution accumulate different dynamical phases and rotate through phase space at different rates. This differential rotation progressively shears and bends the outer regions of the distribution, producing the observed curved tails and transforming the initially Gaussian delocalized state into a distinctly non-Gaussian phase-space distribution.

\section{Broader Implications}

Thermally delocalized motional states provide a practical resource for force and displacement sensing with trapped nanoparticles. For a charged  $234~\mathrm{nm}$-diameter silica particle, placed in electric fields~\cite{cg5v-lpjy}, assuming a mechanical resonance frequency of $2\pi\times55~\mathrm{kHz}$, a damping rate of $2\pi\times200~\mathrm{Hz}$, and a temperature of $7.8~\mathrm{K}$, the thermal force-noise power spectral density is approximately $S_F=2m\gamma k_BT\simeq 4~\mathrm{aN^2/Hz}$~\cite{optomechanicsreview}. Applying $13~\mathrm{dB}$ of momentum squeezing reduces the force-noise amplitude spectral density by a factor of $10^{13/20}\simeq4.47$, corresponding to a reduction of the force-noise power spectral density by a factor of $10^{13/10}\simeq20$. Thus, $S_F^{(\mathrm{sq})}\simeq4/20\simeq0.20~\mathrm{aN^2/Hz}$, or, equivalently, the force-noise amplitude spectral density is reduced from $2~\mathrm{aN/\sqrt{Hz}}$ to approximately $0.45~\mathrm{aN/\sqrt{Hz}}$. This corresponds to a nearly fivefold improvement in amplitude sensitivity over the feedback-cooled thermal state.

Polarization-adaptive delocalization controls the axial trap stiffness without altering optical power which can increase the recoil heating rate at ultra-high vacuum~\cite{Recoil_Jain} [3 times for $\theta_p=88.3^\circ$, as shown in Appendix \ref{appendixF}]. However, the standing-wave geometry provides an additional means of suppressing recoil heating. Near a dark fringe, the recoil rate scales as $k_l^2\langle z^2\rangle$ where $\langle z^2 \rangle$ is the position variance [see Appendix \ref{appendixF}], allowing strong suppression in the Lamb--Dicke regime ($k_l \sqrt{\langle z^2\rangle} \ll 1$) while retaining finite trap curvature. As the delocalization increases, however, recoil heating grows, establishing a trade-off between motional delocalization and recoil heating.

The motional delocalization is strongest along the axial direction, with only a modest $\leq 1\text{ dB}$ radial anti-squeezing observed during the polarization-adaptive protocol due to the accompanying reduction in radial trap stiffness. Conversely, no radial delocalization is observed during dark-fringe adaptive protocols. At an optical node, the radial potential flattens ($U(r) \approx 0$), removing the radial restoring force without creating an inverted radial trap. Because the dark-fringe placement duration ($\tau_D = 1.8\ \mu\text{s}$) is orders of magnitude shorter than the radial period ($T_r \approx 770\ \mu\text{s}$), the particle undergoes negligible thermal drift, leaving the transverse phase-space distribution effectively frozen. To achieve simultaneous multi-mode squeezing, parametric power-modulation protocols~\cite{delocalisation, Kamba2025QuantumSqueezing} can be integrated with polarization- and dark-fringe-adaptive schemes to dynamically modulate trap stiffness across all motional degrees of freedom.

With ongoing improvements in our laboratory’s vacuum technology, we estimate that an average phonon number below unity can be achieved at a pressure of approximately ($10^{-7}~\mathrm{mbar}$) which is demonstrated in hollow-core fibers~\cite{PhysRevA.83.063830, linder}. Beyond delocalization, the sign-insensitive response of the dark fringe provides a route toward measurement-induced spatial superpositions with ground-state cooled fringe-trapped particle. Since the dark fringe lies midway between two adjacent bright fringes, the particle-scattered light intensity follows $\hat{I}_d(\hat{z})\propto\hat{z}^2$ which can enable a quadratic position measurement. Such measurement can then conditionally prepare a coherent spatial Schrödinger-cat state of two symmetric positions $\pm z_m$~\cite{romero_pra_2011}. Subsequent free evolution will cause the two components to overlap and produce matter-wave interference fringes. Therefore, dark fringes can prove to be a vital resource for testing macroscopic quantum physics.

\section{Concluding Remarks} 
In summary, we demonstrated advanced optomechanical control of a levitated nanoparticle, trapped in the interference fringes of an HC-PCF. We achieved simultaneous feedback cooling of axial and radial modes and generated axial delocalization by both extinguishing the fringes and transporting the particle into and out of dark inverted optical fringes. We also demonstrated generation of a non-Gaussian motional state. Our measurements agree closely with a Wiener dynamics-based analytical model capturing deterministic evolution and stochastic thermal fluctuations. These results establish fringe-based optical traps hollow-core fibers as a versatile platform for deterministic control of levitated solid-state particles. We anticipate that our work will open a route toward fiber-based generation and transport of non-classical states of motion for enhanced sensing as well as advanced optomechanics experiments.

\section*{Acknowledgments} 
We acknowledge financial support from the Max Planck Society and the Deutsche Forschungsgemeinschaft (OrbitFlySens: ID-418737652).   
\\
\section*{Author contributions} S. C. and G. W. performed the experiments. P. K. developed the theoretical model and performed the numerical simulations. C. G. and N. J. supervised the project. All authors contributed in writing and editing the paper.  
\\
\section*{Conflict of Interest} The authors declare that they have no conflict of interest.
\\
\section*{Data availability} The datasets generated and/or analysed during the current study are not publicly available while the manuscript is under peer review. They will be deposited in an appropriate public repository and made publicly available upon publication. The repository name and accession number (or DOI) will be provided before publication.

\onecolumngrid
\appendix

%	\section*{Supplementary Text}
%	 This Supplemental Text details: (1) detailed covariance matrix based analysis of the particle's motion under dissipative thermal noise, (2) theoretical and numerical analysis of the motional squeezing in the polarization-adaptive methods (frequency jumps), (3) analytical and numerical investigations on the particle's motional dynamics in the dark harmonic potential, (4) extension of the motional analysis for the anharmonic optical potential, and (5) analytical comparison of the recoil heating rates during polarization-adaptive squeezing and dark-fringe trap based squeezing.
	
%%%%%%%%%%%%%%%% MATERIALS AND METHODS %%%%%%%%%%%%%%%
\section{Materials and Methods}
\label{appendixA}

The polarization-adaptive delocalization is performed using four consecutive steps, repeated over 3000 iterations. 

(1) The particle is initialized in a low-energy motional state at a axial trap frequency of $\Omega_0=2\pi\times 55$~kHz under feedback cooling. The cooling ensures our homodyne-type detection remains in the linear regime. The feedback is then switched off for $20 ~\mu s$.

(2) The standing-wave pattern is transiently suppressed, therefore reducing the axial trap frequency to $\Omega_1=2\pi\times 10.2$~kHz  for a duration of $T_1/4$  where $T_1=2\pi/\Omega_1$. The particle motion gets squeezed during the frequency jump. By choosing the evolution duration correctly in the shallow optical potential~\cite{Kamba2025QuantumSqueezing}, we can maximize the amount of squeezing (see main text).   

(3) The standing-wave trap is restored. 

(4) The feedback is switched on and the system is allowed to reinitialize for $10~\mathrm{ms}$.

(5) The entire sequence is repeated 3000 times. 

For the entire duration of the protocol, the axial position of the particle was monitored in real time using a FPGA-based lock-in amplifier around the axial eigenfrequency, providing continuous positional information through the orthogonal in-phase and quadrature components. The velocity can be reconstructed either by numerical differentiation of the measured position~\cite{vorvosh} or directly from the quadrature component~\cite{romain-dark}; both methods yielded consistent results for us. To account for the finite response of the detection chain, we further use the retrodiction protocol introduced in Refs.~\cite{rossi_retro, northrup_sq, romain-dark}, in which the measured position is used to reconstruct the velocity at an earlier time. A retrodiction test, confirms that, except $1~\rm \mu s$ after the pulse, transients arising from AC coupling do not affect the reconstructed velocity. The resulting position and velocity define the motional phase-space trajectory, and all phase-space data are processed using the same protocol. To correct for slow positional drift~\cite{romain-dark}, the 3000 realizations are divided into 100 groups of 30 consecutive realizations, with each group spanning more than 300 ms. For each group, the mean position is calculated, and the resulting 100 group means are linearly fitted to determine the positional drift. The fitted drift is then subtracted from the mean position of each group before constructing the phase-space distributions. This correction centers the groups at a common position and prevents slow positional drift from artificially increasing the measured position variance and, consequently, from leading to an overestimation of the inferred squeezing.

Here we detail the protocol for squeezing in dark trap which is repeated 3000 times.

(1) Starting from the cold thermal state and keeping the feedback off, the standing-wave pattern is transiently shifted by $\lambda/4 = 266~\mathrm{nm}$, transferring the particle to a dark optical fringe.

(2) The particle evolves in the dark fringe for $1.8~\mu\mathrm{s}$. For longer dark-pulse durations, we observe that the particle becomes axially unstable, as indicated by sharp, irregular excursions in the measured position.

(3) The standing-wave pattern is shifted back, returning the particle to a bright optical fringe and thereby preventing particle loss. The particle subsequently evolves for T/2, where T is the motional oscillation period. 

(4) Afterwards the particle is returned back to the dark fringe where it is allowed to evolve for another 1.8~$\mu s$, following its subsequent return to the bright fringe. One full excursion from $\rm bright\rightarrow dark \rightarrow bright$ fringes is referred to as `dark pulse'. The motional state is measured after returning the particle to the bright fringe after each dark pulse. The velocity was constructed as before using the position data.

(5) The maximum motional squeezing was observed for three consecutive dark pulses while maintaining Gaussian statistics. 

(6) The entire sequence is repeated 3000 times.

(7) Feedback cooling is switched on, and the system is allowed to reinitialize for $10~\mathrm{ms}$.

For the dataset capturing the non-Gaussian motional state using four consecutive dark pulses, 6000 experimental realizations were recorded. For this specific protocol, feedback cooling was reactivated $5\ \mu\text{s}$ after measuring the non-Gaussian state (see Fig.~\ref{nonlinear}) to maintain stable particle trapping at low pressure.

\section{Stochastic thermal evolution}
\label{appendixB}
We consider a spherical, isotropic dielectric particle of mass $m$ confined in a standing-wave optical trap formed by two counterpropagating laser beams. For simplicity, the beams are assumed to have plane-wave like transverse profiles. The exact treatment of the transverse mode profile is discussed in the next subsection \ref{appendixB1}. The beams have identical frequency $\omega_{\ell}$ and equal but opposite wave vectors $\pm k_{\ell}$ along $z$-direction. The positive-frequency part of the electric field is a coherent superposition of $\hat{e}_{L}\mathcal{E}_{0}e^{ik_\ell z}e^{-i\omega_\ell t}$ and $\hat{e}_{R}\mathcal{E}_{0}e^{-ik_\ell z}e^{-i\omega_\ell t}$ $e^{-2i\phi}$ counterpropagating waves, where $\mathcal{E}_0$ denotes the common field amplitude, $\phi(t)$ is the relative phase between the optical fields, $\hat{e}_L$ and $\hat{e}_R$ are the polarization unit vectors of the left- and right-propagating beams, respectively. To eliminate the rapid optical oscillation, we transform to a frame rotating at the optical frequency $\omega_\ell$. In this rotating frame, the total electric field is given by~\cite{Kumar}
\begin{align}
E(z,t)&=\mathcal{E}_{0}\Big[\hat{e}_{L}e^{ik_{\ell}z}+\hat{e}_{R}e^{-ik_{\ell}z}e^{-2i\phi(t)}\Big]\;.\label{eq_electric_field}
\end{align}
This field gives rise to a position- and time-dependent potential energy $U(z,t) = -\alpha' |E(z,t)|^2 / 4$, where $\alpha' \equiv \Re(\alpha)$ is the real part of the particle's complex polarizability. For small deviations from the trap minimum, the optical force is well approximated by a dipole gradient force acting on the particle's center of mass. Assuming the particle remains localized at the potential minimum and ignoring the constant shift, this force takes the form
\begin{align}
F(z)
    =
    -\frac{\partial U(z,t)}{\partial z}
    =
    \frac{\alpha'}{4}
    \frac{\partial |E(z, t)|^2}{\partial z}\approx-\kappa_{\rm trap}\,z\cos[2\phi(t)]\;,\label{eq_optical_force}
\quad \\ \text{with axial stiffness}~~
\kappa_{\rm trap} = 2k_{\ell}^{2}\alpha'|\mathcal{E}_{0}|^{2}\cos(\theta_p)
\end{align}
where $\theta_p$ is is the relative polarization angle defined by $\hat{e}_{L}\cdot\hat{e}_{R} = \cos(\theta_p)$. Recognizing that $\kappa_{\rm trap} = \kappa_o \cos(\theta_p)$ with $\kappa_o = 2k_{\ell}^{2}\alpha'|\mathcal{E}_{0}|^{2} = m\Omega_o^2$, the axial resonant frequency becomes
\begin{align}
\Omega = \Omega_o \sqrt{\cos(\theta_p)}\;.\label{eq_frequency}
\end{align}

Crucially, this reveals that the axial frequency $\Omega$ can be tuned directly by adjusting the relative polarization angle $\theta_p$ between the two beams. Unlike power-based tuning, which requires changing the laser intensity and may introduce recoil heating issues, polarization control offers a fast, in-situ mechanism for modifying the trapping potential without altering the material properties $m$ and $\alpha'$, or the overall laser power. Furthermore, the relative phase of the counterpropagating fields can shift the standing-wave pattern, thereby relocating the particle from bright to dark fringes. Accordingly, at $\phi=0$ the particle experiences a restoring force in the harmonic potential at the bright fringe, while at $\phi=\pi/2$ it experiences a repulsive force in the inverted trap at the dark fringe. For polarization-adaptive delocalization (see Main Text) $\theta_p$ is changed from $0^\circ$ to $88.3^\circ$, therefore reducing the axial frequency from 55 kHz to 10.2 kHz.

\subsection{On-axis intensity and radial stiffness}
\label{appendixB1}

When the relative polarization orientation between the trapping beams $\theta_p$ is switched from $0^\circ$ to $88.3^\circ$ the radial frequency also shifts to 0.92 kHz from 1.3 kHz. Here we explain the reasoning behind it. To account for the transverse fiber mode profile, we use~\cite{Chakraborty:26}
\begin{align}
E(r)=AJ_0\!\left(\frac{2u_{01}r}{d_{\rm core}}\right),
\end{align}
where $A$ is the maximum electric field amplitude at the center of the core, \(d_{\rm core}\) is the core diameter and \(u_{01}=2.4048\) is the first zero of \(J_0\), such that \(J_0(u_{01})=0\). Including this mode profile in Eq.~\eqref{eq_electric_field}, the intensity is
\begin{align}
|E(r,z,t)|^2
&=
|A|^2
J_0^2\!\left(\frac{2u_{01}r}{d_{\rm core}}\right)
\Big[
2+2\cos(\theta_p)
\cos\!\big[2k_\ell z+2\phi(t)\big]
\Big].
\end{align}
At a bright fringe ($z=0,~\phi=0$) and on axis ($r=0$),
\begin{align}
|E|^2_{\mathrm{bright}}(0)
=
2|A|^2[1+\cos(\theta_p)],
\end{align}
so that
\begin{align}
\frac{|E|^2_{\mathrm{bright}}(90^\circ)}
{|E|^2_{\mathrm{bright}}(0^\circ)}
=
\frac{1}{2}.
\end{align}

Near the axis,
\begin{align}
J_0^2\!\left(\frac{2u_{01}r}{d_{\rm core}}\right)
\simeq
1-\frac{2u_{01}^2r^2}{d_{\rm core}^2}.
\end{align}
For a dispersive optical potential, \(U(r)\propto-|E(r)|^2\), giving radial stiffness as
\begin{align}
\kappa_r
\propto
\frac{2u_{01}^2}{d_{\rm core}^2}
|A|^2[1+\cos(\theta_p)].
\end{align}
Thus, assuming the transverse mode profile and each trapping beam intensity are unchanged with polarization,
\begin{align}
\frac{\kappa_r(\theta_p)}{\kappa_r(0)}
&=
\frac{1+\cos(\theta_p)}{2},
\\
\frac{\Omega_r(\theta_p)}{\Omega_r(0)}
&=
\sqrt{\frac{1+\cos(\theta_p)}{2}}.
\label{radial}
\end{align}
For \(\theta_p=88.3^\circ\), Eq.~\eqref{radial} predicts
\begin{align}
\frac{\Omega_r(88.3^\circ)}{\Omega_r(0)}
\simeq0.711,
\end{align}
in agreement with the measured ratio
\begin{align}
\frac{0.92}{1.3}\simeq0.708.
\end{align}

\subsection{Equations of motion}
To capture the stochastic evolution of the particle's center-of-mass motion, we employ the theoretical framework detailed in Ref.~\cite{Kumar,ghosh2023phase}. By setting $\phi=0$, the dynamics is formulated as a system of coupled stochastic difference equations:
\begin{subequations}  
\begin{align}  
dq &= \Omega p \, dt, \\  
dp &= -\gamma p \, dt - \Omega q \, dt + \sqrt{2\gamma n_{\mathrm{th}}} \, dW^{\text{th}}(t)\;.
\end{align}\label{eq_Eqs_Motion}    
\end{subequations}  
The variables $q$ and $p$ appearing above denote the dimensionless position and momentum quadratures, defined respectively as $q = z / q_{\mathrm{zpm}} $ and $ p = p_z / p_{\mathrm{zpm}}$, where $q_{\mathrm{zpm}} = \sqrt{\hbar/(m\Omega)}$ and $p_{\mathrm{zpm}} = \sqrt{m\hbar\Omega}$ are the zero-point position and momentum scales. The stochastic forcing term arises from thermal collisions with surrounding gas molecules, effectively coupling the oscillator to a reservoir at temperature $T_{\text{th}}$ with mean occupancy $n_{\text{th}}$. This interaction manifests as a pressure-dependent friction coefficient $\gamma$ and a Gaussian white-noise process represented by the Wiener increments $dW^{\text{th}}(t)$, which obey $\langle dW^{\text{th}}(t)\rangle = 0$ and $\langle (dW^{\text{th}}(t))^2 \rangle = dt$.

\subsection{Covariance matrix analysis}
\label{Cov_matrix_bright}
To obtain an analytical treatment of the system, we discretize the interval $[0, t]$ into $n$ equal segments of length $dt = t/n$. This procedure transforms the stochastic differential equations (Eqs.~\eqref{eq_Eqs_Motion}) into an equivalent set of recurrence relations, which explicitly account for thermal fluctuations at each discrete timestep. In the underdamped limit 
($\gamma\ll\Omega$), recursive back-substitution in conjunction with a long-time asymptotic analysis yields the steady-state solutions for the position and momentum quadratures. These solutions can be expressed compactly as
\begin{align}
    \mathbf{u}(t) &=e^{-\gamma\tau/2}\mathcal{R}(\Omega,\tau)\mathbf{u}(t_{0})+\mathbf{N}_{\Omega}(t,t_{0})\;, \label{eq_shorthand_quadratures}
\end{align}
where $\mathbf{u}(t)=\begin{bmatrix} q(t), p(t) \end{bmatrix}^{T}$ is the mechanical quadrature vector. Using the particle's axial eigenfrequency $\Omega$ as defined in the Eq.~\eqref{eq_frequency}, the rotation matrix in Eq.~\eqref{eq_shorthand_quadratures} is given by
\begin{align}
   \mathcal{R}(\Omega\tau) &= \begin{bmatrix}
        \cos(\Omega\tau) & \sin(\Omega\tau)\\
        -\sin(\Omega\tau) & \cos(\Omega\tau)
    \end{bmatrix}\;, \label{eq_rotation_matrix}
\end{align}
with $\tau = t - t_{0}$. The diffusion vector in Eq.~\eqref{eq_shorthand_quadratures} is defined as $\mathbf{N}_{\Omega}(t,t_{0}) = \begin{bmatrix} \mathcal{N}^{\Omega}_{q}(t,t_{0}),  \mathcal{N}^{\Omega}_{p}(t,t_{0}) \end{bmatrix}^{T}$, whose components are
\begin{align}
  \mathcal{N}^{\Omega}_{q}(t,t_{0})&=  -\frac{i}{2}\sqrt{2\gamma n_{\mathrm{th}}}\lim_{n\rightarrow\infty}\sum_{j=0}^{n-1}(\lambda_{-}^{j}-\lambda_{+}^{j})dW^{\mathrm{th}}_{n-j-1}\;, \label{eq_position_diffusion}\\
  \mathcal{N}^{\Omega}_{p}(t,t_{0})&= -\frac{i}{4\Omega}\sqrt{2\gamma n_{\mathrm{th}}}\lim_{n\rightarrow\infty}\sum_{j=0}^{n-1}(\gamma_{+}\lambda_{+}^{j}-\gamma_{-}\lambda_{-}^{j})dW^{\mathrm{th}}_{n-j-1}\;, \label{eq_momentum_diffusion}
\end{align}
where $\lambda_{\pm}=1-\gamma_{\pm}dt/2$ are the eigenvalues of the evolution matrix, with $\gamma_{\pm}=\gamma\pm\bar{\Omega}$ and $\bar{\Omega}=\sqrt{\gamma^{2}-4\Omega^{2}}$. 

In the presence of thermal noise Wiener increments, the above equations yield the phase-space distribution, whose covariance matrix is given by
\begin{align}
    V(t) &=e^{-\gamma\tau}\mathcal{R}(\Omega\tau)V(0)\mathcal{R}^{T}(\Omega\tau)+\mathcal{Q}(t,t_{0}). \label{eq_Cov_Mat}
\end{align}
The first term arises from the deterministic dynamics and accounts for the rotation and relaxation of the initial distribution, characterized by the covariance
\begin{align}
    V(0) = \langle \mathbf{u}(0)\mathbf{u}^{T}(0)\rangle = n_{\mathrm{th}}\,\mathcal{I}, \label{eq_Cov_initial}
\end{align}
where \(\mathcal{I}\) is the \(2\times2\) identity matrix. The second term,
\begin{align}
    \mathcal{Q}(t,t_{0}) &= \langle \mathbf{N}_{\Omega}(t,t_{0}) \mathbf{N}^{T}_{\Omega}(t,t_{0}) \rangle= 
       \begin{bmatrix}
         \left\langle\left(\mathcal{N}^{\Omega}_{q}(t,t_{0})\right)^{2}\right\rangle & 
         \left\langle\mathcal{N}^{\Omega}_{q}(t,t_{0})\mathcal{N}^{\Omega}_{p}(t,t_{0})\right\rangle \\[8pt]
         \left\langle\mathcal{N}^{\Omega}_{q}(t,t_{0})\mathcal{N}^{\Omega}_{p}(t,t_{0})\right\rangle & 
         \left\langle\left(\mathcal{N}^{\Omega}_{p}(t,t_{0})\right)^{2}\right\rangle
       \end{bmatrix}\;, \label{eq:Diffusion_Matrix}
\end{align}
represents the contribution of thermal noise diffusion to the covariance evolution. Here, we have invoked the symmetry of the covariance matrix, which implies that the cross-correlation terms satisfy $\left\langle\mathcal{N}^{\Omega}_{q}(t,t_{0})\mathcal{N}^{\Omega}_{p}(t,t_{0})\right\rangle=\left\langle\mathcal{N}^{\Omega}_{p}(t,t_{0})\mathcal{N}^{\Omega}_{q}(t,t_{0})\right\rangle$. Using the Wiener increment property $\langle dW_{n-j-1}^{\mathrm{th}}dW_{n-j'-1}^{\mathrm{th}}\rangle=\delta_{jj'}dt$ and evaluating the resulting geometric sums, the diffusion contribution to the covariance matrix  in the underdamped regime becomes
\begin{align}
    \mathcal{Q}(t,t_{0}) &= n_{\mathrm{th}}(1-e^{-\gamma\tau})\mathcal{I}\;.\label{eq_Difussion_contribution}
\end{align}
Finally, using the rotation matrix [Eq. \eqref{eq_rotation_matrix}] and Eq. \eqref{eq_Difussion_contribution}, the simplified form of the covariance matrix is
\begin{align}
    V(t)&=n_{\mathrm{th}}\mathcal{I}\;.
\end{align}

\section{Analysis of squeezing via sudden frequency quench}
\label{appendixC}
We now analyze the mechanical squeezing generated by rapid modulation of the trapping potential. The analysis is formulated in terms of the covariance matrix of the motional quadratures, allowing the evolution of the principal variances and the resulting squeezing to be determined directly. We first consider a single sudden frequency jump, which establishes the basic mechanism of transient squeezing. The frequency jump is experimentally performed by rapidly rotating the relative polarization orientation ($\theta_p$) between the trapping beams from $0^\circ$ to $88.3^\circ$ \big[see Eq.~\eqref{eq_frequency}\big]. We then extend the analysis to a finite square pulse, where the frequency is switched for a controlled duration before being restored to its initial value. 

\subsection{Sudden frequency jump}
We first consider the elementary case of a sudden change in the relative polarization of the trapping beams leading to sudden quench in the mechanical frequency~\cite{Janszky1986}. Such rapid frequency jumps provide one of the simplest and most efficient methods for generating mechanical squeezing. The abrupt change in the oscillator
frequency modifies the relative scaling of the position and momentum quadratures, resulting in a redistribution of fluctuations between the two conjugate variables while preserving the total phase-space area. Immediately after the
quench, one quadrature exhibits reduced fluctuations below its equilibrium value, whereas the conjugate quadrature experiences an increased variance, producing a squeezed state. In the absence of dissipation, the squeezed state undergoes periodic rotation in phase space at the new oscillation frequency while maintaining constant principal variances. In realistic systems, however, damping and thermal noise gradually restore thermal equilibrium, causing the squeezing to decay over time.

To visualize mechanical squeezing via a rapid frequency jump, we model the process as an instantaneous transition from $\Omega_{0}$ to $\Omega_{1}$ at the switching time $t_{1}$, where $\Omega_{0}\neq\Omega_{1}$. The resulting time-dependent frequency is thus given by the piecewise function
\begin{align}
\Omega(t) = 
\begin{cases} 
\Omega_0, & 0 \le t < t_1, \\[6pt]
\Omega_1, & t \ge t_1,
\end{cases}\;,\label{eq_freq_jump}
\end{align}

\begin{figure}[ht!]
\centering
    \includegraphics[width=0.6\linewidth]{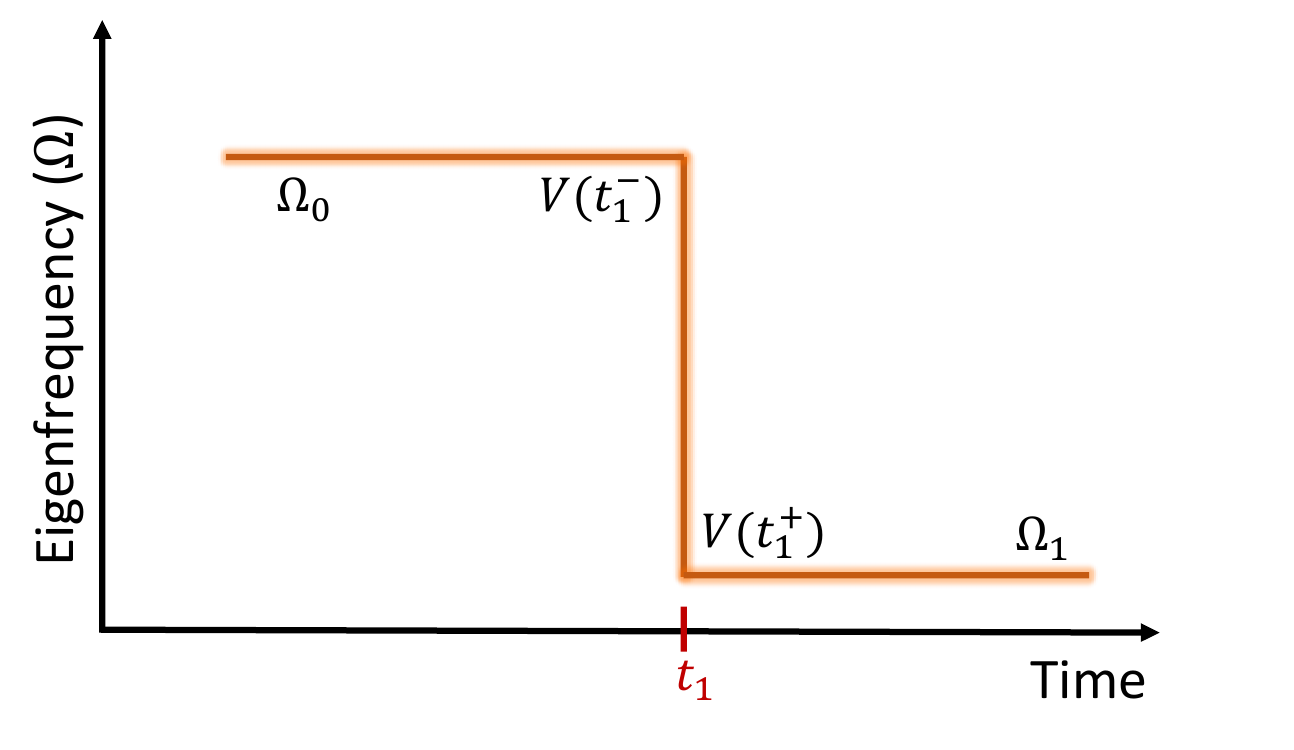}
    \caption{Instantaneous axial eigenfrequency $\Omega$ as a function of time. The oscillator operates at its initial frequency $\Omega_0$ for $0 \le t < t_1$, and at the switching instant $t_1$ it undergoes an abrupt step to $\Omega_1$, remaining constant thereafter. The plot illustrates the idealized step discontinuity described by the piecewise frequency model. $V(t_{1}^{-})$ and $V(t_{1}^{-})$ are the covariance matrices of the particle motion just before and after the frequency jump, respectively.}
    \label{figS1}
\end{figure}

In Fig. \ref{figS1}, we depict a specific case of such a jump at a switching time $t=t_{1}$.

\subsubsection{Just before jump}
To establish the covariance matrix  immediately before the frequency quench, we evaluate Eq.~\eqref{eq_Cov_Mat} for the pre-quench evolution $0\leq t<t_{1}$. At $t=t_{1}^{-}$, the covariance matrix is
\begin{align}
    V(t_{1}^{-})
    =
    e^{-\gamma t_{1}}
    \mathcal{R}(\Omega_{0}t_{1})
    V(0)
    \mathcal{R}^{T}(\Omega_{0}t_{1})
    +
    \mathcal{Q}(t_{1},0)\;.
    \label{justbefore}
\end{align}

For an initially thermal state, $V(0)=n_{\mathrm{th}}^{\Omega_{0}}\mathcal{I}$, where $n_{\mathrm{th}}^{\Omega_{0}}=k_{B}T_{\mathrm{th}}/\hbar\Omega_{0}$ is the thermal occupation corresponding to the initial mechanical frequency $\Omega_{0}$. Substituting the rotation matrix from Eq.~\eqref{eq_rotation_matrix} and the corresponding diffusion contribution from Eq.~\eqref{eq_Difussion_contribution}, the dissipative evolution preserves the thermal covariance, yielding
\begin{align}
    V(t_{1}^{-})=n_{\mathrm{th}}^{\Omega_{0}}\mathcal{I}\;.
    \label{eq_Cov_before_jump}
\end{align}

Thus, immediately before the frequency jump, the dimensionless position and momentum quadratures have equal variances and zero cross-correlation. The motional state is therefore isotropic, corresponding to a circular phase-space distribution. This provides a well-defined thermal initial state for the subsequent nonadiabatic evolution following the frequency quench.

\paragraph*{Dimensional covariance matrix before the frequency jump:}

To relate the dimensionless covariance matrix to the experimentally relevant position and momentum fluctuations, we introduce the dimensional phase-space vector
\begin{align}
    \mathbf{u}_{z}(t)
    &=
    \begin{bmatrix}
        z(t)\\
        p_z(t)
    \end{bmatrix}
    =
    \mathcal{D}(\Omega)\mathbf{u}(t)\;,\label{eq_uz_to_u}
\end{align}
where $\mathbf{u}(t)=\begin{bmatrix}
    q(t),p(t)
\end{bmatrix}^{T}$ is the mechanical dimensionless quadrature vector and the scaling matrix is
\begin{align}
    \mathcal{D}(\Omega)
    &=
    \begin{bmatrix}
        \sqrt{\dfrac{\hbar}{m\Omega}} & 0\\
        0 & \sqrt{m\hbar\Omega}
    \end{bmatrix}\;.\label{eq_scaling_matrix}
\end{align}
Evidently, above scaling matrix is symmetric i.e. $\mathcal{D}^{T}(\Omega)=\mathcal{D}(\Omega)$. The corresponding inverse transformation is
\begin{align}
    \mathbf{u}(t)
    &=
    \mathcal{D}^{-1}(\Omega)\mathbf{u}_{z}(t)\;,\label{eq_u_to_uz}
\end{align}
with
\begin{align}
    \mathcal{D}^{-1}(\Omega)
    &=
    \begin{bmatrix}
        \sqrt{\dfrac{m\Omega}{\hbar}} & 0\\
        0 & \dfrac{1}{\sqrt{m\hbar\Omega}}
    \end{bmatrix}\;.\label{eq_inv_scaling_matrix}
\end{align}

Under this transformation, the covariance matrix in dimensional variables is
\begin{align}
    V_z(t)
    &=
    \left\langle
        \mathbf{u}_z(t)\mathbf{u}_z^{T}(t)
    \right\rangle
    =
    \mathcal{D}(\Omega)
    V(t)
    \mathcal{D}^{T}(\Omega)\;.\label{eq_Cov_mat_z}
\end{align}

Immediately before the frequency jump, the dimensionless covariance matrix is given by Eq. \eqref{eq_Cov_before_jump}. Thus, 
evaluating the transformation at the pre-quench frequency $\Omega_0$ gives
\begin{align}
    V_z(t_1^{-})
    &=\mathcal{D}(\Omega_{0})V(t_{1}^{-})\mathcal{D}^{T}(\Omega_{0})=
    n_{\mathrm{th}}^{\Omega_0}
    \begin{bmatrix}
        \dfrac{\hbar}{m\Omega_0} & 0\\
        0 & m\hbar\Omega_0
    \end{bmatrix}.
\end{align}

Thus, the isotropic thermal state in the dimensionless phase space corresponds to the expected thermal position and momentum variances in physical units, with their respective scales set by the mechanical frequency $\Omega_0$.

\subsubsection{At the frequency jump}

At $t=t_{1}$, the trapping frequency is suddenly quenched from $\Omega_{0}$ to $\Omega_{1}$. Because the quench is assumed to be instantaneous, the dimensional position and momentum remain unchanged during the switching process. Consequently, the dimensional phase-space vector and its covariance matrix are continuous across the jump,
\begin{align}
    \mathbf{u}_{z}(t_{1}^{+})&=\mathbf{u}_{z}(t_{1}^{-}),
    \qquad
    V_{z}(t_{1}^{+})=V_{z}(t_{1}^{-})\;.
\end{align}

Although the physical position and momentum remain unchanged, their dimensionless representations change because their natural scaling depends on the mechanical frequency. This change in scaling generates the squeezing transformation associated with the frequency quench.

\subsubsection{Just after the frequency jump}

Using the dimensional-to-dimensionless transformation at the new frequency $\Omega_{1}$, together with the continuity of $\mathbf{u}_{z}$, the post-quench dimensionless quadratures are related to their pre-quench values by
\begin{align}
    \mathbf{u}(t_{1}^{+})
    &=
    \mathcal{D}^{-1}(\Omega_{1})
    \mathcal{D}(\Omega_{0})
    \mathbf{u}(t_{1}^{-})
    =
    \mathcal{S}_{10}\mathbf{u}(t_{1}^{-})\;,
\end{align}
where
\begin{align}
    \mathcal{S}_{10}
    &=
    \begin{bmatrix}
        \sqrt{\dfrac{\Omega_{1}}{\Omega_{0}}} & 0\\
        0 & \sqrt{\dfrac{\Omega_{0}}{\Omega_{1}}}
    \end{bmatrix}\;.
\end{align}

This diagonal transformation stretches one dimensionless quadrature while compressing its conjugate. Introducing the squeezing parameter
\begin{align}
    r
    &=
    \frac{1}{2}\ln\left(\frac{\Omega_{1}}{\Omega_{0}}\right)\;,\label{eq_squeezing_parameter}
\end{align}
the transformation takes the standard squeezing form
\begin{align}
    \mathcal{S}_{10}
    &=
    \begin{bmatrix}
        e^{r} & 0\\
        0 & e^{-r}
    \end{bmatrix}\;.\label{eq_squeezing_matrix}
\end{align}
Thus,
\begin{align}
    q(t_{1}^{+})&=e^{r}q(t_{1}^{-}),
    \qquad
    p(t_{1}^{+})=e^{-r}p(t_{1}^{-}).
\end{align}

The corresponding covariance matrix immediately after the jump is
\begin{align}
    V(t_{1}^{+})
    &=
    \mathcal{S}_{10}
    V(t_{1}^{-})
    \mathcal{S}_{10}^{T}
    =
    n_{\mathrm{th}}^{\Omega_{0}}
    \begin{bmatrix}
        e^{2r} & 0\\
        0 & e^{-2r}
    \end{bmatrix}\;.\label{eq_Cov_mat_immediatley_after_jump}
\end{align}

Thus, the sudden frequency quench transforms the initially isotropic thermal covariance into an anisotropic Gaussian covariance, with one quadrature amplified and its conjugate compressed. This provides the initial squeezed state for the subsequent evolution at the new mechanical frequency.

\subsubsection{At a later time after the jump}
For times $t>t_{1}$, the covariance matrix evolves under the post-quench mechanical frequency $\Omega_{1}$. Defining the elapsed time after the jump as $\tau_{1}=t-t_{1}$, we have
\begin{align}
    V(t)&=\langle \mathbf{u}(t)\mathbf{u}^{T}(t)\rangle=e^{-\gamma\tau_{1}}\mathcal{R}(\Omega_{1}\tau_{1})V(t_{1}^{+})\mathcal{R}^{T}(\Omega_{1}\tau_{1})+\mathcal{Q}_{2}(t,t_{1})\;,\label{eq_Cov_mat_after_jump}
\end{align}
where $\mathcal{Q}_{2}(t,t_{1})=n_{\mathrm{th}}^{\Omega_{1}}\Big(1-e^{-\gamma \tau_{1}}\Big)\mathcal{I}$. Expressing the covariance matrix as
\begin{align}
    V(t)&=\begin{bmatrix}
        V_{qq}(t) & V_{qp}(t)\\
        V_{pq}(t) & V_{pp}(t)
    \end{bmatrix}\;.\label{eq_Cov_mat_form}
\end{align}
Here, the diagonal elements, $V_{qq}(t)$ and $V_{pp}(t)$, describe the variances of the position and momentum quadratures, while the off-diagonal elements $V_{qp}(t)=V_{pq}(t)$ quantify the correlation between them. 
Substituting the covariance matrix immediately after the jump and evaluating the matrix products gives the covariance matrix elements for $t>t_{1}$
\begin{align}
    V_{qq}(t)\Big\vert_{t\geq t_{1}}
    &=
    e^{-\gamma \tau_{1}}n_{\mathrm{th}}^{\Omega_{0}}
    \left[
        \cosh(2r)
        +\cos(2\Omega_{1}\tau_{1})\sinh(2r)
    \right]
    +n_{\mathrm{th}}^{\Omega_{1}}
    \left(1-e^{-\gamma \tau_{1}}\right),
    \label{eq_Vqq_after_jump}
\end{align}
\begin{align}
    V_{pp}(t)\Big\vert_{t\geq t_{1}}
    &=
    e^{-\gamma \tau_{1}}n_{\mathrm{th}}^{\Omega_{0}}
    \left[
        \cosh(2r)
        -\cos(2\Omega_{1}\tau_{1})\sinh(2r)
    \right]
    +n_{\mathrm{th}}^{\Omega_{1}}
    \left(1-e^{-\gamma \tau_{1}}\right),
    \label{eq_Vpp_after_jump}
\end{align}
\begin{align}
    V_{qp}(t)\Big\vert_{t\geq t_{1}}=V_{pq}(t)\Big\vert_{t\geq t_{1}}
    &=
    -e^{-\gamma \tau_{1}}n_{\mathrm{th}}^{\Omega_{0}}
    \sin(2\Omega_{1}\tau_{1})\sinh(2r).
    \label{eq_Vqp_after_jump}
\end{align}
Eqs. (\ref{eq_Vqq_after_jump})-(\ref{eq_Vqp_after_jump}) provide the complete analytical time evolution of the covariance matrix following the sudden frequency jump. Since the phase-space distribution remains Gaussian throughout the evolution, the covariance matrix completely characterizes its shape and orientation. The main features of these solutions are summarized below.
\begin{enumerate}
    \item \textbf{Evolution of the quadrature variances:} The diagonal covariance elements describe the fluctuations of the position and momentum quadratures. Immediately after the sudden frequency jump, the initially isotropic thermal distribution is transformed into an anisotropic state, resulting in unequal fluctuations along the two quadratures. This anisotropy is the direct signature of mechanical squeezing generated by the sudden change in the oscillator frequency.
    
    \item \textbf{Periodic exchange of quadrature fluctuations:} As evident from Eqs.~(\ref{eq_Vqq_after_jump}) and (\ref{eq_Vpp_after_jump}), the position and momentum variances oscillate at twice the post-jump oscillator frequency with opposite phase, indicating a periodic exchange of fluctuations between the two quadratures. The oscillation amplitude is governed by the squeezing parameter $r$, while the exponential factor describes the gradual suppression of these oscillations due to damping.
    
    \item \textbf{Emergence of position--momentum correlation:} Unlike the equilibrium thermal state, where the covariance matrix is diagonal, the sudden frequency jump generates a finite off-diagonal covariance, $V_{qp}=V_{pq}$. This correlation arises because the squeezed phase-space distribution is generally rotated with respect to the position and momentum axes, causing the two quadratures to become statistically correlated.
    
    \item \textbf{Rotation of the squeezing axis:} As shown by Eq.~(\ref{eq_Vqp_after_jump}), the position--momentum correlation oscillates sinusoidally at twice the post-jump oscillator frequency while decaying exponentially with the damping rate. Physically, the non-zero covariance indicates that the principal axes of the covariance ellipse are no longer aligned with the coordinate axes. Instead, the squeezing axis continuously rotates in phase space until the correlation is gradually erased by dissipation.
    
    \item \textbf{Relaxation to thermal equilibrium:} In the long-time limit ($\tau_1\rightarrow\infty$), the exponentially decaying terms vanish, yielding
    \begin{align}
    V_{qq}=V_{pp}=n_{\mathrm{th}}^{\Omega_{1}}, \qquad
    V_{qp}=0\;.\label{eq_Cov_elements_t_infinity}
    \end{align}
    Consequently, the covariance matrix becomes diagonal and isotropic, corresponding to the thermal equilibrium state associated with the final oscillator frequency $\Omega_{1}$. All transient squeezing, anisotropy, and quadrature correlations generated by the sudden frequency jump are therefore completely removed by the interaction with the thermal environment.
\end{enumerate}

\subsubsection{Squeezing characterization}

To quantify the anisotropy generated by the sudden frequency jump, we diagonalize the covariance matrix. For the covariance matrix $V(t)$, the two principal variances are
\begin{align}
\Lambda_{\mathrm{max,min}}(t)
=
\frac{V_{qq}(t)+V_{pp}(t)}{2}
\pm
\frac{1}{2}
\sqrt{
\left[V_{qq}(t)-V_{pp}(t)\right]^2
+4V_{qp}^2(t)
}.
\label{eq_eigenvalues}
\end{align}

\begin{figure}[ht!]
\centering
    \includegraphics[width=0.5\linewidth]{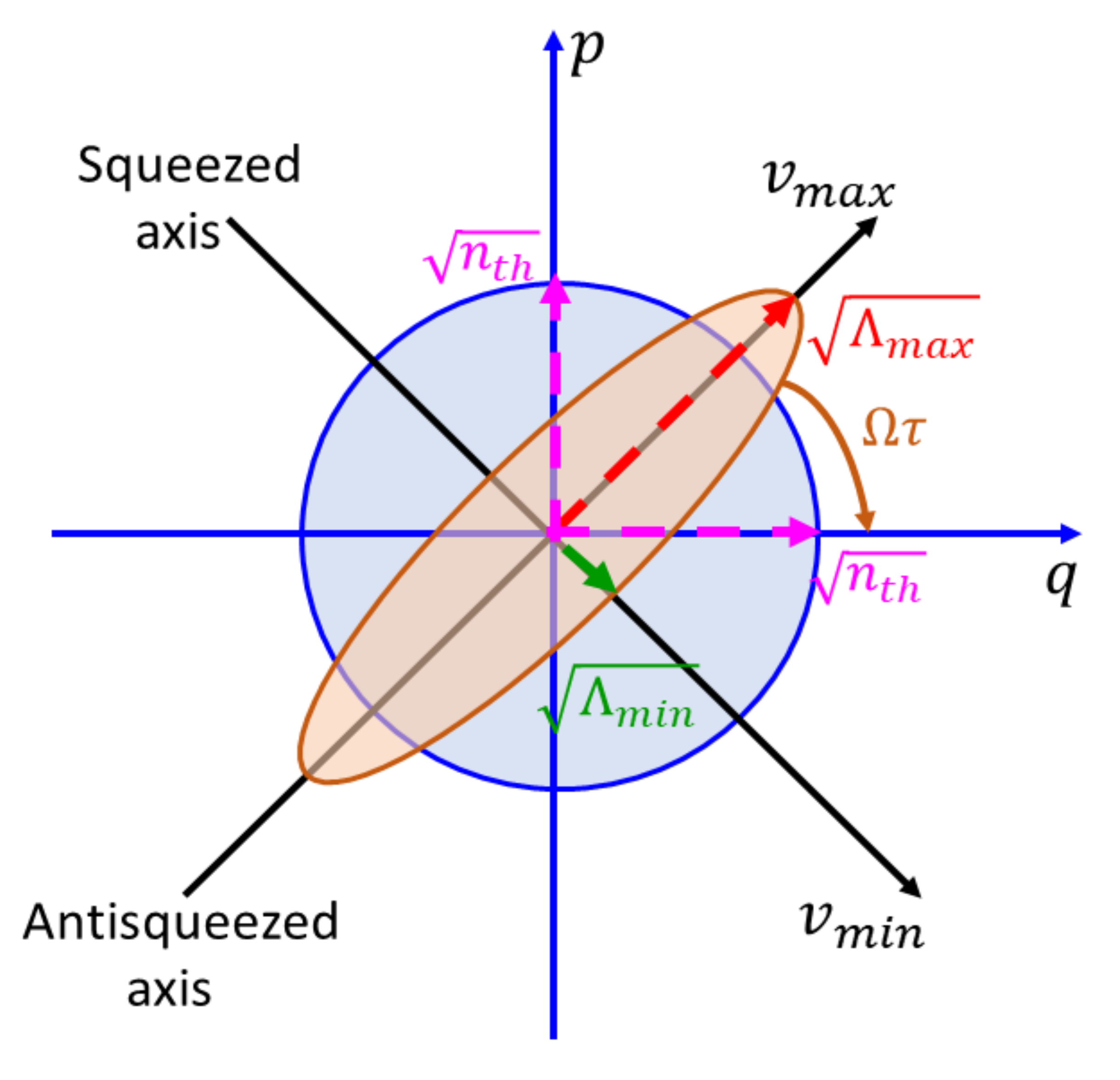}
    \caption{\textbf{Phase-space schematic under a single frequency quench.}
        The schematic illustrates the transformation of the motional state in phase space. Initially, the oscillator is in an isotropic thermal state, represented by a circle of radius $\sqrt{n_{\rm th}}$, where $n_{\rm th}$ is the mean thermal occupancy. An instantaneous frequency quench at $t_1$ deforms this circle into a squeezed ellipse with principal axes along the squeezed and antisqueezed quadratures, having variances $\Lambda_{\min}$ and $\Lambda_{\max}$, respectively. The semi-minor and semi-major axis lengths are $\sqrt{\Lambda_{\min}}$ and $\sqrt{\Lambda_{\max}}$. Subsequently, the ellipse rotates clockwise in phase space at the trap frequency $\Omega$, as described by the harmonic rotation $\mathcal{R}(\Omega\tau)$. Damping and thermal diffusion gradually restore the isotropic thermal state at long times.
    }
    \label{Fig_SM_phase_space}
\end{figure}

Here, $\Lambda_{\mathrm{max}}$ and $\Lambda_{\rm min}$ denote the maximum and minimum variances, respectively, along the principal axes of the phase-space covariance ellipse, as shown schematically in Fig. \ref{Fig_SM_phase_space}. Substitution of the covariance elements derived above gives
\begin{align}
\Lambda_{\max}(t)
&=
e^{-\gamma\tau_1}
n_{\mathrm{th}}^{\Omega_0}e^{2|r|}
+
n_{\mathrm{th}}^{\Omega_1}
\left(1-e^{-\gamma\tau_1}\right),
\label{eq_max_eigenvalue}\\
\Lambda_{\min}(t)
&=
e^{-\gamma\tau_1}
n_{\mathrm{th}}^{\Omega_0}e^{-2|r|}
+
n_{\mathrm{th}}^{\Omega_1}
\left(1-e^{-\gamma\tau_1}\right).
\label{eq_min_eigenvalue}
\end{align}

The two eigenvalues are the principal variances of the Gaussian phase-space distribution, with corresponding covariance-ellipse semi-axis lengths proportional to $\sqrt{\Lambda_{\min}}$ and $\sqrt{\Lambda_{\max}}$. The associated eigenvectors determine the orientation of these principal axes. The coherent contribution retains the anisotropy generated by the frequency jump, while damping attenuates this contribution and thermal diffusion drives both principal variances toward the same equilibrium value.

The normalized eigenvectors are
\begin{align}
\mathbf{v}_{\max}
&=
\begin{bmatrix}
\cos(\Omega_1\tau_1)\\
-\sin(\Omega_1\tau_1)
\end{bmatrix},
\qquad
\mathbf{v}_{\min}
=
\begin{bmatrix}
\sin(\Omega_1\tau_1)\\
\cos(\Omega_1\tau_1)
\end{bmatrix}.
\label{eq_eigenvectors}
\end{align}

Thus, the eigenvector associated with $\Lambda_{\min}$ defines the instantaneous squeezed quadrature, whereas that associated with $\Lambda_{\max}$ defines the anti-squeezed quadrature. Their explicit dependence on $\Omega_1\tau_1$ shows that the principal axes rotate in phase space at the post-quench mechanical frequency. Thermal diffusion changes the magnitudes of the principal variances but does not alter these eigenvectors.

To characterize the anti-squeezing of the position quadrature in decibel units, we express the ratio of minimum variance to the initial thermal occupation $n_{\rm th}^{\Omega_{0}}$ on a logarithmic scale as
\begin{align}
\mathcal{S}(t)
=
-10\log_{10}
\left(
\frac{\Lambda_{\min}(t)}
{n_{\rm th}^{\Omega_{0}}}
\right).
\label{eq_squeezing_dB}
\end{align}

\subsubsection*{Pre-quench dynamics: $0\leq t<t_1$}

For completeness, before the frequency jump the oscillator evolves at the initial frequency $\Omega_0$. Starting from the thermal covariance $V(0)=n_{\mathrm{th}}^{\Omega_0}\mathcal{I}$, the rotational evolution does not alter the isotropic covariance, while the diffusion contribution maintains thermal equilibrium. Consequently, throughout the interval $0\leq t<t_1$,
\begin{align}
V(t)=n_{\mathrm{th}}^{\Omega_0}\mathcal{I}.
\end{align}

The two eigenvalues are therefore degenerate,
\begin{align}
\Lambda_{\max}(t)=\Lambda_{\min}(t)=n_{\mathrm{th}}^{\Omega_0},
\qquad 0\leq t<t_1.
\end{align}

Thus, the phase-space distribution remains circular and has no preferred principal direction before the frequency jump. The corresponding squeezing is $\mathcal{S}(t)=0$. 

\subsubsection*{Limiting cases after the jump}

The squeezing expression has two particularly transparent limiting cases.

\begin{enumerate}

\item \textbf{Immediately after the frequency jump, $\tau_1=0$:} The diffusion contribution has not yet accumulated, so the principal variances are
$\Lambda_{\max}(0)=n_{\mathrm{th}}^{\Omega_0}e^{2|r|}$ and
$\Lambda_{\min}(0)=n_{\mathrm{th}}^{\Omega_0}e^{-2|r|}$.
And the squeezing is
\begin{align}
\mathcal{S}_{\max}\Big\vert_{\tau_{1}=0}=10\Bigg\vert\log_{10}\Big(\frac{\Omega_{1}}{\Omega_{0}}\Big)\Bigg\vert\;.
\end{align}
This is the maximum squeezing immediately following the quench.

\item \textbf{Long-time limit, $\tau_1\rightarrow\infty$:} The contribution retaining the memory of the initial state is exponentially suppressed, while thermal diffusion drives both principal variances toward the equilibrium value $n_{\mathrm{th}}^{\Omega_1}$. Hence,
$\Lambda_{\max}=\Lambda_{\min}=n_{\mathrm{th}}^{\Omega_1}$, giving
$\mathcal{S}(\infty)=0$.
The covariance ellipse consequently becomes circular, corresponding to thermal equilibrium at the final mechanical frequency.

\end{enumerate}

Overall, the analytical result captures the competition between quench-induced coherent anisotropy and thermalization: the former is exponentially attenuated by damping, while the latter drives both principal variances toward the same thermal equilibrium value. The resulting anisotropy consequently decreases from its maximum immediately after the frequency jump to zero at long times.

\subsection{Squeezing analysis for a square pulse}
\begin{figure}[ht!]
    \centering
     \includegraphics[width=0.72\linewidth]{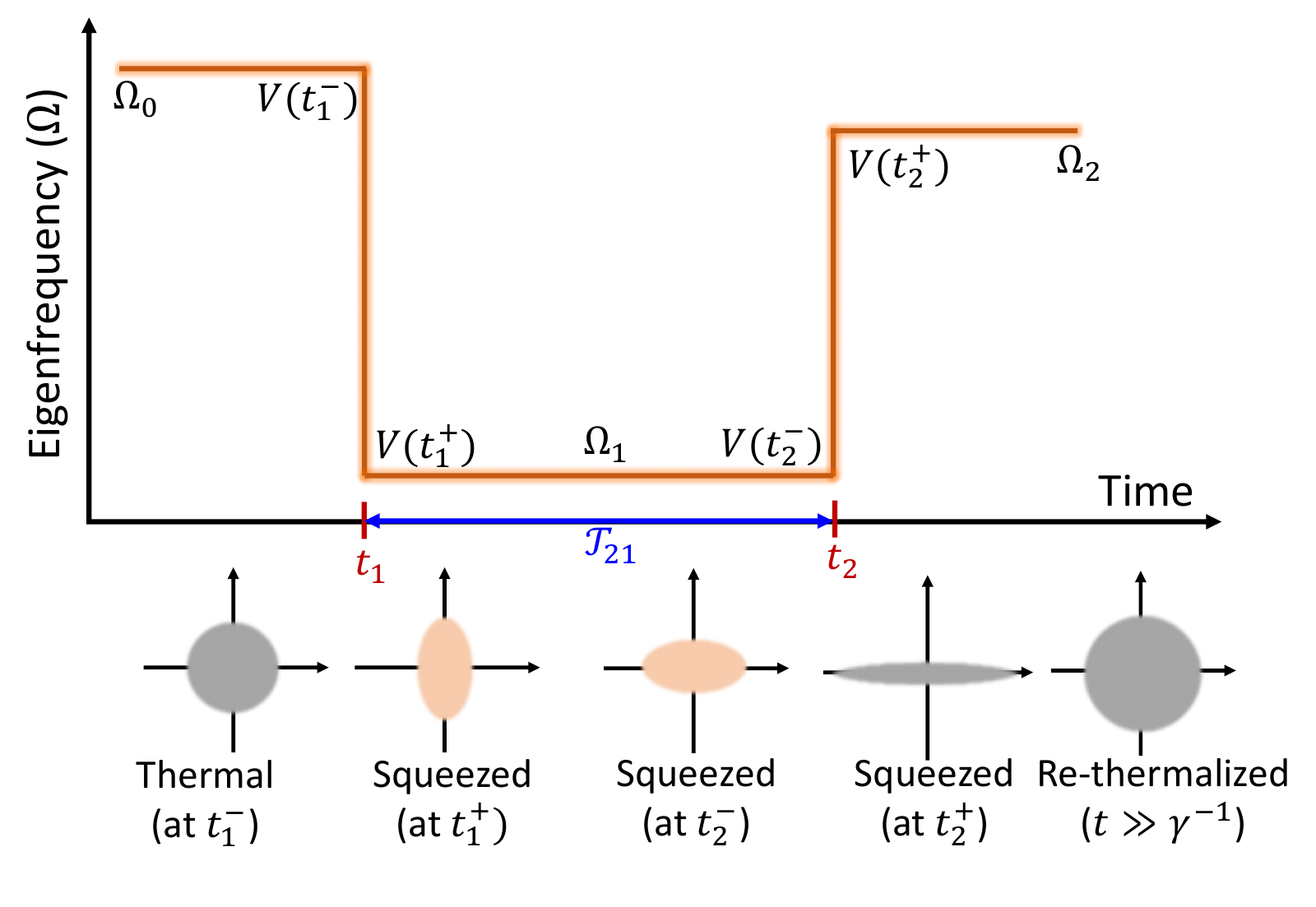}
    \caption{
        \textbf{Squeezing via a square frequency pulse.}
        The schematic depicts the time-dependent trap frequency (top) and the corresponding phase-space distributions (bottom). Two rapid polarization rotations at $t_1$ and $t_2$ produce a square-pulse frequency modulation, $\Omega_0 \to \Omega_1 \to \Omega_2$. Experimentally, $\Omega_2=\Omega_0$ as considered in Eq.~\eqref{omega2isomega0}.
        The covariance matrices $V(t_{1}^{-})$ and $V(t_{1}^{+})$ denote the states immediately before and after the first quench, respectively; similarly, $V(t_{2}^{-})$ and $V(t_{2}^{+})$ denote the states immediately before and after the second quench. The first quench squeezes the initial thermal state; the quarter-period evolution at $\Omega_1$ rotates the squeezed ellipse by $\pi/2$; and the second quench, acting on this rotated state, further enhances the squeezing. In the long-time limit, damping and thermal diffusion erase the anisotropy, and the system rethermalizes to an isotropic thermal distribution. The protocol demonstrates deterministic, feedback-free control of thermal noise.
    }
    \label{Fig_SM_square_pulse}
\end{figure}

Above we have described a detailed squeezing analysis for a single frequency jump. Now, we will extend the analysis when there are two successive frequency quenches
\begin{align}
	\Omega_0
	\rightarrow
	\Omega_1
	\rightarrow
	\Omega_2,
	\end{align}
with switching times $t_{1}$ and $t_{2}$, respectively. Fig. \ref{Fig_SM_square_pulse} depicts the schematic of such a pulse.

Importantly, we need to drive the covariance matrix $V(t)$ for $t>t_{2}$ and obtain its eigenvalues to characterize squeezing. 
\subsubsection{First frequency quench $\Omega_{0}\rightarrow\Omega_{1}$}
Let us start from the initial covariance matrix 
\begin{align}
    V(0) &= n_{\mathrm{th}}^{\Omega_{0}}\mathcal{I}\;,
\end{align}
and defining $\tau_{1}=t_{1}$, we can write its evolution upto a time just below the first quench
\begin{align}
    V(t_{1}^{-})&=e^{-\gamma \tau_{1}}\mathcal{R}(\Omega_{0}\tau_{1})V(0)\mathcal{R}^{T}(\Omega_{0}\tau_{1})+n_{\mathrm{th}}^{\Omega_{0}}(1-e^{-\gamma \tau_{1}})\mathcal{I}=n_{\mathrm{th}}^{\Omega_{0}}\mathcal{I}\;.
\end{align}
Note that just before the first quench, the covariance matrix is diagonal with variances $n_{\mathrm{th}}^{\Omega_{0}}$. This shows that the covariance matrix remains proportional to the identity. Consequently, the position and momentum quadratures remain uncorrelated and the thermal distribution stays circular in phase space despite the deterministic phase-space rotation. The situation will be drastically different at the second frequency quench where the covariance matrix will involve off-diagonal contributions.

Now, the covariance matrix just above the first quench is goverened by
\begin{align}
    V(t_{1}^{+})&=S_{10}(r_{10})V(t_{1}^{-})S^{T}_{10}(r_{10})\;,
\end{align}
where the squeezing matrix and the associated parameter with the first quench are
\begin{align}
    S_{10}&=\begin{bmatrix}
        e^{r_{10}} & 0\\
        0 & e^{-r_{10}}
    \end{bmatrix},\quad r_{10}=\frac{1}{2}\log_{e}\Big(\frac{\Omega_{1}}{\Omega_{0}}\Big)\;.
\end{align}
Then the covariance matrix for any time $t$ over the interval $[t_{1},t_{2})$ and with the notation $\tau_{2}=t-t_{1}$ is
\begin{align}
    V(t)\Big\vert_{t_{1}\leq t<t_{2}}&=e^{-\gamma \tau_{2}}\mathcal{R}(\Omega_{1}\tau_{2})V(t_{1}^{+})\mathcal{R}^{T}(\Omega_{1}\tau_{2})+n_{\mathrm{th}}^{\Omega_{1}}(1-e^{-\gamma \tau_{2}})\mathcal{I}\;.
\end{align}

\subsubsection{Second frequency quench $\Omega_{1}\rightarrow\Omega_{2}$}
Just before the second quench, we can write the covariance matrix by defining $\mathcal{T}_{21}=t_{2}-t_{1}$
\begin{align}
    V(t_{2}^{-})&=e^{-\gamma \mathcal{T}_{21}}\mathcal{R}(\Omega_{1}\mathcal{T}_{21})V(t_{1}^{+})\mathcal{R}^{T}(\Omega_{1}\mathcal{T}_{21})+n_{\mathrm{th}}^{\Omega_{1}}(1-e^{-\gamma \mathcal{T}_{21}})\mathcal{I}\;.
\end{align}
In general, $V(t_{2}^{-})$ contains both unequal diagonal elements and finite off-diagonal correlations, indicating that the phase-space distribution is simultaneously squeezed and rotated with respect to the canonical quadratures. Just after the second frequency quench, the symplectic transformation connects the solution to the jump before it
\begin{align}
    V(t_{2}^{+})&=S_{21}(r_{21})V(t_{2}^{-})S^{T}_{21}(r_{21})\;,
\end{align}
where the updated squeezing matrix and associated parameter are
\begin{align}
    S_{21}&=\begin{bmatrix}
        e^{r_{21}} & 0\\
        0 & e^{-r_{21}}
    \end{bmatrix},\quad r_{21}=\frac{1}{2}\log_{e}\Big(\frac{\Omega_{2}}{\Omega_{1}}\Big)\;.
\end{align}
Unlike the first quench, the second symplectic transformation acts on a covariance matrix that is already squeezed and rotated. Consequently, the resulting covariance cannot, in general, be diagonalized analytically by inspection, and the principal variances must be obtained from the eigenvalues of $V(t)$.

Finally, the covariance matrix for $t>t_{2}$ is governed by
\begin{align}
    V(t)\Big\vert_{t_{2}\leq t}&=e^{-\gamma \tau_{3}}\mathcal{R}(\Omega_{2}\tau_{3})V(t_{2}^{+})\mathcal{R}^{T}(\Omega_{2}\tau_{3})+n_{\mathrm{th}}^{\Omega_{2}}(1-e^{-\gamma \tau_{3}})\mathcal{I}\;,
\end{align}
where $\tau_{3}=t-t_{2}$.

\subsubsection{Covariance matrix elements}
Now let us find the matrix elements of above covariance matrix. For this, just before the second frequency quench, the covariance matrix can be expressed in terms of its elements as
\begin{align}
V(t_{2}^{-})
=
\begin{bmatrix}
V_{qq}(\mathcal{T}_{21}) &
V_{qp}(\mathcal{T}_{21})
\\
V_{qp}(\mathcal{T}_{21}) &
V_{pp}(\mathcal{T}_{21})
\end{bmatrix}\;,\label{eq_V_t2neg}
\end{align}
where the covariance elements are obtained from the dissipative evolution over the interval
$t_{1}\leq t<t_{2}$.

Applying the second symplectic transformation gives
\begin{align}
V(t_{2}^{+})
&=
S_{21}V(t_{2}^{-})S_{21}^{T}=
\begin{bmatrix}
e^{2r_{21}}V_{qq}(\mathcal{T}_{21})
&
V_{qp}(\mathcal{T}_{21})
\\
V_{qp}(\mathcal{T}_{21})
&
e^{-2r_{21}}V_{pp}(\mathcal{T}_{21})
\end{bmatrix}.
\end{align}

For $t\geq t_{2}$ the covariance matrix evolves as
\begin{align}
V(t)
=
e^{-\gamma\tau_{3}}\,
\mathcal{R}(\Omega_{2},\tau_{3})
V(t_{2}^{+})
\mathcal{R}^{T}(\Omega_{2},\tau_{3})
+
n_{\mathrm{th}}^{\Omega_{2}}\left(1-e^{-\gamma\tau_{3}}\right)
\mathcal{I}.
\end{align}

The individual covariance elements are therefore
\begin{align}
V_{qq}(t)
&=
e^{-\gamma\tau_{3}}
\Big[
e^{2r_{21}}V_{qq}(\mathcal{T}_{21})
\cos^{2}(\Omega_{2}\tau_{3})+
e^{-2r_{21}}V_{pp}(\mathcal{T}_{21})
\sin^{2}(\Omega_{2}\tau_{3})
+
V_{qp}(\mathcal{T}_{21})
\sin(2\Omega_{2}\tau_{3})
\Big]+
n_{\mathrm{th}}^{\Omega_{2}}
\left(1-e^{-\gamma\tau_{3}}\right)\;,
\label{eq_Vqq_square_pulse}
\\[1ex]
V_{pp}(t)
&=
e^{-\gamma\tau_{3}}
\Big[
e^{2r_{21}}V_{qq}(\mathcal{T}_{21})
\sin^{2}(\Omega_{2}\tau_{3})
+ e^{-2r_{21}}V_{pp}(\mathcal{T}_{21})
\cos^{2}(\Omega_{2}\tau_{3})
-
V_{qp}(\mathcal{T}_{21})
\sin(2\Omega_{2}\tau_{3})
\Big]+
n_{\mathrm{th}}^{\Omega_{2}}
\left(1-e^{-\gamma\tau_{3}}\right)\;,
\label{eq_Vpp_square_pulse}
\\[1ex]
V_{qp}(t)
&=
\frac{e^{-\gamma\tau_{3}}}{2}
\Big[
\Big(
e^{-2r_{21}}V_{pp}(\mathcal{T}_{21})
-
e^{2r_{21}}V_{qq}(\mathcal{T}_{21})
\Big)
\sin(2\Omega_{2}\tau_{3})
+
2V_{qp}(\mathcal{T}_{21})
\cos(2\Omega_{2}\tau_{3})
\Big]\;.
\label{eq_Vqp_square_pulse}
\end{align}

The principal variances of the covariance ellipse are obtained from the eigenvalues of
$V(t)$ and can be obtained by using Eq. \eqref{eq_eigenvalues}. The covariance elements
$V_{qq}(\mathcal{T}_{21})$,
$V_{pp}(\mathcal{T}_{21})$,
and
$V_{qp}(\mathcal{T}_{21})$ in Eqs. \eqref{eq_Vqq_square_pulse}-\eqref{eq_Vqp_square_pulse} can be explicitly expressed as
\begin{align}
V_{qq}(\mathcal{T}_{21})
&=
e^{-\gamma \mathcal{T}_{21}}
n_{\mathrm{th}}^{\Omega_{0}}
\Big[
\cosh(2r_{10})
+
\cos(2\Omega_{1}\mathcal{T}_{21})
\sinh(2r_{10})
\Big]+
n_{\mathrm{th}}^{\Omega_{1}}
\left(
1-e^{-\gamma \mathcal{T}_{21}}
\right)\;,\label{eq_Vqq_T21}
\\
V_{pp}(\mathcal{T}_{21})
&=
e^{-\gamma \mathcal{T}_{21}}
n_{\mathrm{th}}^{\Omega_{0}}
\Big[
\cosh(2r_{10})
-
\cos(2\Omega_{1}\mathcal{T}_{21})
\sinh(2r_{10})
\Big]+
n_{\mathrm{th}}^{\Omega_{1}}
\left(
1-e^{-\gamma \mathcal{T}_{21}}
\right)\;,\label{eq_Vpp_T21}\\
V_{qp}(\mathcal{T}_{21})
&=
-
e^{-\gamma \mathcal{T}_{21}}
n_{\mathrm{th}}^{\Omega_{0}}
\sin(2\Omega_{1}\mathcal{T}_{21})
\sinh(2r_{10})\;.\label{eq_Vqp_T21}
\end{align}

\subsubsection{Pulse duration $\mathcal{T}_{21}=nT_1/4$}

The general square-pulse solution derived above is valid for an arbitrary pulse duration
$\mathcal{T}_{21}=t_{2}-t_{1}$.
In general, the covariance matrix immediately before the second quench contains
non-vanishing position--momentum correlations (Eq. \eqref{eq_Vqp_T21}),
which originate from the continuous phase-space rotation of the squeezed thermal state during its evolution under the trap frequency $\Omega_{1}$.
Consequently, the principal axes of the covariance ellipse are generally rotated with respect to the position and momentum quadratures, leading to the more complicated expressions derived above.

A particularly important simplification occurs when the pulse duration satisfies
\begin{align}
\mathcal{T}_{21}
=
\frac{n\pi}{2\Omega_{1}}=\frac{n T_{1}}{4},
\qquad
n=0,1,2,\ldots
\label{eq:principal_axis_condition}
\end{align}
At these stroboscopic times, the covariance ellipse realigns with the position and momentum axes, giving $V_{qp}(\mathcal{T}_{21})=0$. Physically, this corresponds to the absence of instantaneous position--momentum correlations even though the state remains squeezed. In other words, the squeezing axes coincide with the canonical quadratures, so that the covariance matrix becomes diagonal immediately before the second frequency quench.

Applying Eq.~(\ref{eq:principal_axis_condition}) to the covariance matrix immediately before the second quench, as defined in Eq.~\eqref{eq_V_t2neg}, yields a diagonal matrix,
\begin{align}
V(t_{2}^{-})
=
\begin{bmatrix}
V_{qq}(\mathcal{T}_{21}) & 0\\
0 & V_{pp}(\mathcal{T}_{21})
\end{bmatrix}\;.
\end{align}
Under this condition, the matrix elements in Eqs.~\eqref{eq_Vqq_T21}–\eqref{eq_Vpp_T21} simplify to:
\begin{align}
V_{qq}(\mathcal{T}_{21})
&=
e^{-\gamma\mathcal{T}_{21}}
n_{\mathrm{th}}^{\Omega_{0}}
e^{\pm2r_{10}}
+
n_{\mathrm{th}}^{\Omega_{1}}
\left(
1-e^{-\gamma\mathcal{T}_{21}}
\right),
\\
V_{pp}(\mathcal{T}_{21})
&=
e^{-\gamma\mathcal{T}_{21}}
n_{\mathrm{th}}^{\Omega_{0}}
e^{\mp2r_{10}}
+
n_{\mathrm{th}}^{\Omega_{1}}
\left(
1-e^{-\gamma\mathcal{T}_{21}}
\right),
\end{align}
where the upper (lower) sign corresponds to even (odd) values of $n$.

Since the covariance matrix is diagonal at $t=t_{2}^{-}$, its principal axes coincide with the canonical position and momentum quadratures. The second frequency quench therefore simply rescales the two principal variances according to
\begin{align}
V(t_{2}^{+})
&=
S_{21}
V(t_{2}^{-})
S_{21}^{T}=
\begin{bmatrix}
e^{2r_{21}}V_{qq}(\mathcal{T}_{21}) & 0\\
0 & e^{-2r_{21}}V_{pp}(\mathcal{T}_{21})
\end{bmatrix},
\end{align}

The subsequent dissipative evolution acts independently on each principal variance. Consequently, the eigenvalues [Eq.~\eqref{eq_eigenvalues}] of the covariance matrix  evaluate to:

\begin{subequations}
\label{eq:square_pulse_simplified}
\begin{align}
\Lambda_{\mathrm{max}}(t)
&=
e^{-\gamma\tau_{3}}
e^{2r_{21}}
\Big[
e^{-\gamma\mathcal{T}_{21}}
n_{\mathrm{th}}^{\Omega_{0}}
e^{\pm2r_{10}}
+
n_{\mathrm{th}}^{\Omega_{1}}
\left(
1-e^{-\gamma\mathcal{T}_{21}}
\right)
\Big]
+
n_{\mathrm{th}}^{\Omega_{2}}
\left(
1-e^{-\gamma\tau_{3}}
\right)\;,\\
\Lambda_{\mathrm{min}}(t)
&=
e^{-\gamma\tau_{3}}
e^{-2r_{21}}
\Big[
e^{-\gamma\mathcal{T}_{21}}
n_{\mathrm{th}}^{\Omega_{0}}
e^{\mp 2r_{10}}
+
n_{\mathrm{th}}^{\Omega_{1}}
\left(
1-e^{-\gamma\mathcal{T}_{21}}
\right)
\Big]
+
n_{\mathrm{th}}^{\Omega_{2}}
\left(
1-e^{-\gamma\tau_{3}}
\right)\;.
\end{align}\label{eq:square_pulse_simplified}
\end{subequations}

Representing a direct extension of the single-quench results [Eqs.~\eqref{eq_min_eigenvalue}--\eqref{eq_max_eigenvalue}], Eq.~(\ref{eq:square_pulse_simplified}) describes a square-pulse protocol under the principal-axis revival condition. Experimentally, this regime is realized by synchronizing the second frequency jump with integer multiples of a quarter oscillation period at the intermediate trapping frequency $\Omega_{1}$. Furthermore, the equation naturally separates the dynamics into three distinct contributions: (i) the initially squeezed thermal state prepared by the first quench and damped throughout the pulse duration; (ii) the thermal fluctuations accumulated during evolution under $\Omega_{1}$ that are subsequently transformed by the second quench; and (iii) the post-pulse relaxation toward thermal equilibrium at the final trapping frequency $\Omega_{2}$.

\subsubsection{Return pulse: $\Omega_2=\Omega_0$}
A special case of significant experimental utility is the return-pulse protocol, wherein the trap frequency is restored to its initial value following a finite duration,
\begin{equation}
\Omega_0 \rightarrow \Omega_1 \rightarrow \Omega_0.
\label{omega2isomega0}
\end{equation}
Under these conditions, setting $r_{21} = -r_{10}$ yields the principal variances for even $n$:
\begin{subequations}
\begin{align}
\Lambda_{\mathrm{max}}^{\mathrm{even}}(t)
&=
e^{-\gamma\tau_3} \left[ e^{-\gamma\mathcal{T}_{21}} n_{\rm th}^{\Omega_0} + n_{\rm th}^{\Omega_1} \left(1-e^{-\gamma\mathcal{T}_{21}}\right) e^{-2|r_{10}|}\right] + n_{\rm th}^{\Omega_0} \left(1-e^{-\gamma\tau_3}\right),
\label{eq:max_even}\\
\Lambda_{\mathrm{min}}^{\mathrm{even}}(t)
&=
e^{-\gamma\tau_3} \left[ e^{-\gamma\mathcal{T}_{21}} n_{\rm th}^{\Omega_0} + n_{\rm th}^{\Omega_1} \left(1-e^{-\gamma\mathcal{T}_{21}}\right) e^{2|r_{10}|}\right] + n_{\rm th}^{\Omega_0} \left(1-e^{-\gamma\tau_3}\right).
\label{eq:min_even}
\end{align}
\end{subequations}
For odd $n$, the corresponding expressions reduce to
\begin{subequations}
\label{eq_eig_odd_m}
\begin{align}
\Lambda_{\mathrm{max}}^{\mathrm{odd}}(t)
&=
e^{-\gamma\tau_3} \left[ e^{-\gamma\mathcal{T}_{21}} n_{\rm th}^{\Omega_0} e^{4|r_{10}|} + n_{\rm th}^{\Omega_1} \left(1-e^{-\gamma\mathcal{T}_{21}}\right) e^{2|r_{10}|}\right] + n_{\rm th}^{\Omega_0} \left(1-e^{-\gamma\tau_3}\right),
\label{eq:max_odd}\\
\Lambda_{\mathrm{min}}^{\mathrm{odd}}(t)
&=
e^{-\gamma\tau_3} \left[ e^{-\gamma\mathcal{T}_{21}} n_{\rm th}^{\Omega_0} e^{-4|r_{10}|} + n_{\rm th}^{\Omega_1} \left(1-e^{-\gamma\mathcal{T}_{21}}\right) e^{-2|r_{10}|}\right] + n_{\rm th}^{\Omega_0} \left(1-e^{-\gamma\tau_3}\right).
\label{eq:min_odd}
\end{align}
\end{subequations}
Note that for the experimental choices $\Omega_{0}/2\pi = 55\text{ kHz}$ and $\Omega_{1}/2\pi = 10.2\text{ kHz}$, the squeezing parameter is negative [$r_{10} = \frac{1}{2}\ln(\Omega_{1}/\Omega_{0}) < 0$], confirming the physical ordering $\Lambda_{\mathrm{min}}^{\mathrm{odd}}(t) < \Lambda_{\mathrm{max}}^{\mathrm{odd}}(t)$.

Finally, the squeezing level expressed in decibels follows directly from Eq.~\eqref{eq_squeezing_dB}.

\subsubsection{Limiting cases}

\paragraph*{(i) Immediately after the second quench.}

At $t=t_2^+$, (i.e., $\tau_3=0$), the eigenvalues reduce to the following explicit forms. For even $n$,
\begin{align}
\Lambda_{\max}^{\mathrm{even}}\Big\vert_{\tau_{3}=0}
&=
e^{-\gamma\mathcal{T}_{21}}
n_{\mathrm{th}}^{\Omega_0}
+
n_{\mathrm{th}}^{\Omega_1}
\left(1-e^{-\gamma\mathcal{T}_{21}}\right)e^{-2|r_{10}|},
\\
\Lambda_{\min}^{\mathrm{even}}\Big\vert_{\tau_{3}=0}
&=
e^{-\gamma\mathcal{T}_{21}}
n_{\mathrm{th}}^{\Omega_0}
+
n_{\mathrm{th}}^{\Omega_1}
\left(1-e^{-\gamma\mathcal{T}_{21}}\right)e^{2|r_{10}|}.
\end{align}

For odd $n$,
\begin{align}
\Lambda_{\max}^{\mathrm{odd}}\Big\vert_{\tau_{3}=0}
&=
e^{-\gamma\mathcal{T}_{21}}
n_{\mathrm{th}}^{\Omega_0}e^{4|r_{10}|}
+
n_{\mathrm{th}}^{\Omega_1}
\left(1-e^{-\gamma\mathcal{T}_{21}}\right)e^{2|r_{10}|},
\\
\Lambda_{\min}^{\mathrm{odd}}\Big\vert_{\tau_{3}=0}
&=
e^{-\gamma\mathcal{T}_{21}}
n_{\mathrm{th}}^{\Omega_0}e^{-4|r_{10}|}
+
n_{\mathrm{th}}^{\Omega_1}
\left(1-e^{-\gamma\mathcal{T}_{21}}\right)e^{-2|r_{10}|}.
\end{align}

These expressions highlight that the squeezing magnitude is directly governed by the phase accrued between the two quenches.

\paragraph*{(ii) Vanishing pulse duration, $\mathcal{T}_{21}\rightarrow0$.}

In this limit, for even $n$, $\Lambda_{\max}^{\mathrm{even}}
=
\Lambda_{\min}^{\mathrm{even}}
=
n_{\mathrm{th}}^{\Omega_0}$,
so that following Eq.~\eqref{eq_squeezing_dB}, the squeezing in dB scale becomes $\mathcal{S}_{\mathrm{even}}\Big\vert_{\tau_{3}=0}^{\mathcal{T}_{21}\rightarrow 0}
\rightarrow0.$
Thus, when the return quench occurs without allowing the oscillator to evolve at $\Omega_1$, the two inverse symplectic transformations cancel.

For odd $n$, the extremal eigenvalues of the phase-space distribution for $\tau_{3}=0$ and $\mathcal{T}_{21}\rightarrow 0$ reduces to $\Lambda_{\max}^{\mathrm{odd}}\vert_{\tau_{3}=0}^{\mathcal{T}_{21}\rightarrow 0}
=
n_{\mathrm{th}}e^{4|r_{10}|}$ and $\Lambda_{\min}^{\mathrm{odd}}\vert_{\tau_{3}=0}^{\mathcal{T}_{21}\rightarrow 0}
=
n_{\mathrm{th}}e^{-4|r_{10}|}$. Using the squeezing parameter $r_{10}=(1/2)\ln(\Omega_{1}/\Omega_{0})$ 
yields
\begin{align}
    \Lambda_{\max}^{\mathrm{odd}}\Big\vert_{\tau_{3}=0}^{\mathcal{T}_{21}\rightarrow 0}
=
n_{\mathrm{th}}\exp\left[2\left|\ln\left(\frac{\Omega_1}{\Omega_0}\right)\right|\right], \qquad
\Lambda_{\min}^{\mathrm{odd}}\Big\vert_{\tau_{3}=0}^{\mathcal{T}_{21}\rightarrow 0}
=
n_{\mathrm{th}} \exp\left[-2\left|\ln\left(\frac{\Omega_1}{\Omega_0}\right)\right|\right]\;.
\end{align}
The amount of squeezing in dB scale for odd $n$ then becomes
\begin{align}
\mathcal{S}_{\mathrm{odd}}\Big\vert_{\tau_{3}=0}^{\mathcal{T}_{21}\rightarrow 0}
=
20\Bigg\vert\log_{10}\left(\frac{\Omega_1}{\Omega_0}\right)\Bigg\vert.
\end{align}

\subsubsection{Numerical results}

\begin{figure}[ht!]
    \centering
     \includegraphics[width=\linewidth]{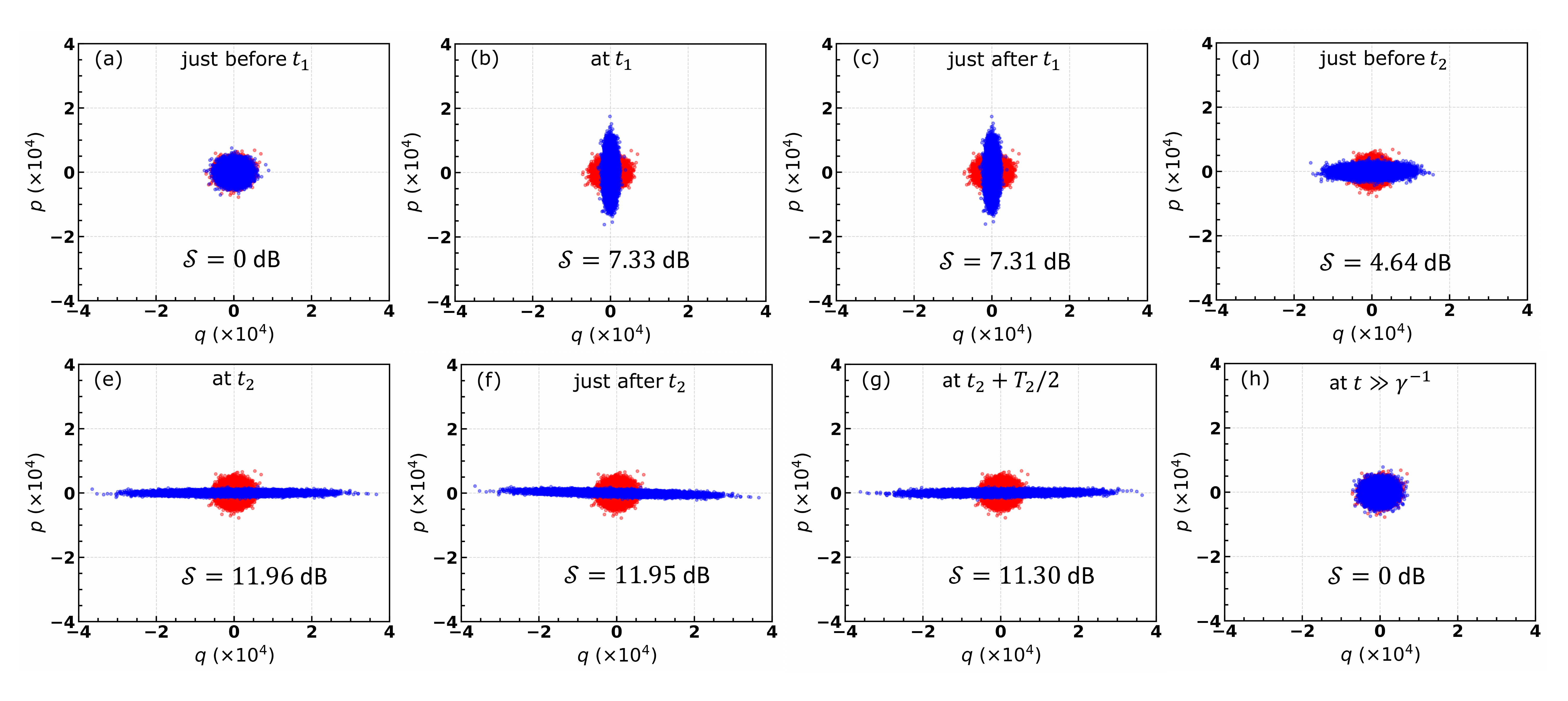}
    \caption{
    \textbf{Numerical simulation of transient squeezing dynamics under a square-pulse protocol.} 
    Phase-space distributions of the motional state at key times during the $\Omega_0 \to \Omega_1 \to \Omega_0$ frequency-quench sequence. (a) Initial isotropic thermal state ($\mathcal{S}=0$~dB). (b,c) The first quench at $t_1$ instantaneously generates momentum squeezing ($\mathcal{S} \approx 7.3$~dB), which subsequently rotates coherently at $\Omega_1$.  (d) Just before $t_2$, the ellipse has rotated by $\pi/2$, reducing squeezing to $\mathcal{S} = 4.64$~dB. 
    (e,f) The return quench at $t_2$ restores $\Omega_0$ and amplifies the anisotropy along the position quadrature, yielding peak squeezing $\mathcal{S} \approx 11.9$~dB. 
    (g) Subsequent rotation at $\Omega_0$ preserves the squeezed axis ($\mathcal{S} = 11.30$~dB). (h) Long-time evolution ($t \gg \gamma^{-1}$) restores the isotropic thermal distribution ($\mathcal{S}=0$~dB) via damping and thermal diffusion. 
    Axes show dimensionless position $q$ and momentum $p$ (scaled by $10^4$). The snapshots are generated from 20,000 trajectories, with initial conditions drawn randomly from a Boltzmann distribution describing a cold thermal state with occupation number $n_{\rm th}$. The oscillator parameters are $\Omega_0/2\pi = 55$ kHz, $\Omega_1/2\pi = 10.2$ kHz, and $\gamma/2\pi = 200$ Hz, with corresponding periods $T_0 = 2\pi/\Omega_0$ and $T_1 = 2\pi/\Omega_1$. The frequency quenches occur at $t_1 = 0.1 T_0$ and $t_2 = t_1 + T_1/4$, where $\mathcal{T}_{21} = T_1/4$ is the pulse duration. The CoM temperature is maintained at $7.8$ K.
}\label{Fig_SM_square_pulse_numerical}
\end{figure}

To reveal squeezing for a polarization square pulse frequency quench protocol, we solve Eqs.~\eqref{eq_Eqs_Motion} numerically via an integrator~\cite{PhysRevA.45.600} with step size $dt = 4 \times 10^{-5}/\gamma$, ensuring stability. The Wiener increments are taken as $dW_{\rm th}(t) = \sqrt{dt} \, \mathcal{N}(0, 1)$, where $\mathcal{N}(0, 1)$ is the standard normal distribution.

Fig.~\ref{Fig_SM_square_pulse_numerical} shows the numerically simulated phase-space evolution of the motional state throughout the square-pulse protocol. Before the first quench, the oscillator occupies an isotropic thermal distribution with zero squeezing (a). At the switching instant $t_1$, the instantaneous frequency jump deforms the state into a squeezed ellipse, yielding $\mathcal{S} = 7.33$~dB (b) and $\mathcal{S} = 7.31$~dB immediately after (c). During the intermediate interval $\mathcal{T}_{21} = T_1/4$, the ellipse rotates coherently at $\Omega_1$ by $\pi/2$, misaligning the squeezing axis with the laboratory frame and reducing the measured value to 4.64~dB just before the second quench (d). At $t_2$, the return quench restores the original trap frequency and applies the inverse symplectic transformation. Immediately after this second quench, the operation amplifies the anisotropy, producing peak transient squeezing of $\mathcal{S} = 11.96$~dB (e) and $\mathcal{S} = 11.95$~dB (f). Throughout the entire temporal evolution, the numerically computed squeezing values agree excellently with the analytical covariance matrix predictions given by Eqs. \eqref{eq:min_odd} and \eqref{eq_squeezing_dB}. Subsequent evolution at $\Omega_0$ preserves the squeezed orientation (g), while long-time dynamics ($t \gg \gamma^{-1}$) gradually erase the anisotropy through damping and thermal diffusion, returning the system to the isotropic thermal state (h).

\section{Squeezing via inverted dark optical potential}
\label{appendixD}
We consider the axial dynamics of the nanoparticle after rapid translation of the standing-wave optical potential from a stable bright fringe to a neighboring dark fringe. The relative optical phase determines the position of the standing-wave pattern along the fiber axis and therefore provides a means of rapidly switching the particle between the two types of trapping regions. On the timescale of the phase switch, which is short compared with the mechanical evolution, the particle’s phase-space coordinates can be taken to remain unchanged. The subsequent dynamics are consequently determined by the local curvature of the optical potential at the dark fringe.

In the vicinity of a stable bright fringe, the axial potential is locally harmonic and the conservative force is restoring. At a dark fringe, the local curvature is inverted, such that the conservative force changes sign while the damping and stochastic forces retain the same form. Accordingly, the coupled finite-difference equations used to describe the axial motion in the stable trap can be transformed to the dark-fringe configuration by reversing the sign of the conservative force term by setting $\phi=\pi/2$ in Eq.~\eqref{eq_optical_force}. In dimensionless quadratures, the resulting equations are 
\begin{align}
    dq &=\Omega pdt\;,\label{eq_dark_diff_q}\\
    dp &= -\gamma p dt +\Omega qdt+\sqrt{2\gamma n_{\mathrm{th}}}dW^{\mathrm{th}}\;.\label{eq_dark_diff_p}
\end{align}
Following the same machinery as introduced in Sec.~\ref{Cov_matrix_bright}, we obtain the solution by mechanical quadratures for underdamped oscillators in an inverted potential
as follows:
\begin{align}
q(t)
&= e^{-\gamma t/2}\cosh(\Omega t)q_0
+ e^{-\gamma t/2}\sinh(\Omega t)p_0
+
\frac{1}{2}\sqrt{2\gamma n_{\mathrm{th}}}
\lim_{n\rightarrow\infty}\sum_{j=0}^{n-1}
\left[\lambda_u^j-\lambda_d^j\right]
dW^{\mathrm{th}}_{n-j-1}\;,
\label{eq_dark_position}\\
p(t)
&= e^{-\gamma t/2}\sinh(\Omega t)q_0
+ e^{-\gamma t/2}\cosh(\Omega t)p_0
+
\frac{1}{2\Omega}\sqrt{2\gamma n_{\mathrm{th}}}
\lim_{n\rightarrow\infty}\sum_{j=0}^{n-1}
\left[\gamma_u\lambda_u^j+\gamma_d\lambda_d^j\right]
dW^{\mathrm{th}}_{n-j-1}\;.
\label{eq_dark_momentum}
\end{align}
The solution naturally separates into a deterministic evolution of the initial quadratures and a stochastic contribution arising from thermal fluctuations. The deterministic part in Eqs. \eqref{eq_dark_position}-\eqref{eq_dark_momentum} follows a hyperbolic propagator whose amplitude decays exponentially, a characteristic of the inverted harmonic potential. In contrast to the oscillatory  evolution of a stable trap, the hyperbolic functions describe exponential separation of the phase-space trajectories. This behavior can be understood in terms of the underlying unstable and decaying eigenmodes, characterized by $\lambda_u=1+\gamma_u dt$ and $\lambda_d=1-\gamma_d dt$, with $\gamma_u=(\bar{\Omega}_D-\gamma)/2$ and $\gamma_d=(\bar{\Omega}_D+\gamma)/2$, where $\bar{\Omega}_D=\sqrt{\gamma^{2}+4\Omega^{2}}$. The unstable mode amplifies exponentially at a rate $\gamma_u$, whereas the decaying mode is attenuated at a rate $\gamma_d$; their combined evolution produces the hyperbolic deterministic dynamics appearing in the solution above. The final terms in both equations account for the thermal diffusion accumulated throughout this unstable evolution.

\subsection{Covariance matrix analysis}
The stochastic solution above provides a direct route to the full statistical description of the nanoparticle’s axial motion in inverted dark potential. We therefore characterize the evolution of the motional state through its covariance matrix, which captures both the deterministic hyperbolic transformation of the initial fluctuations and the diffusion accumulated from thermal noise. This framework allows the separate contributions of the inverted dynamics and stochastic forces to be identified explicitly.

To make the subsequent covariance analysis compact, we express the solutions of Eqs.~\eqref{eq_dark_position} and \eqref{eq_dark_momentum} in matrix form as
\begin{align}
    \mathbf{u}_{D}(t)
    =
    e^{-\gamma t/2}
    \mathcal{O}_{D}(\Omega t)
    \mathbf{u}_{D}(0)
    +
    \mathbf{N}_{D}(t,t_{0})\;,
\end{align}
where
\begin{align}
    \mathbf{u}_{D}(t)
    &=
    \begin{bmatrix}
        q(t)\\
        p(t)
    \end{bmatrix},
    \qquad
    \mathcal{O}_{D}(\Omega t)
    =
    \begin{bmatrix}
        \cosh(\Omega t) & \sinh(\Omega t)\\
        \sinh(\Omega t) & \cosh(\Omega t)
    \end{bmatrix}\;.\label{eq_dark_deterministic_sol}
\end{align}
The matrix $\mathcal{O}_{D}(\Omega t)$ is the hyperbolic propagator of the inverted harmonic potential. It describes the deterministic phase-space evolution in the dark fringe, with the hyperbolic functions reflecting the exponential stretching and contraction of the two conjugate dynamical directions. The vector
\begin{align}
    \mathbf{N}_{D}(t,t_{0})
    =
    \begin{bmatrix}
        \mathcal{N}_{D}^{q}(t,t_{0})\\
        \mathcal{N}_{D}^{p}(t,t_{0})
    \end{bmatrix}\;,
\end{align}
collects the stochastic contribution accumulated during the dark-fringe evolution. Its components are
\begin{align}
    \mathcal{N}_{D}^{q}(t,t_{0})
    &=
    \frac{1}{2}\sqrt{2\gamma n_{\mathrm{th}}}
    \lim_{n\rightarrow\infty}
    \sum_{j=0}^{n-1}
    \left[
        \lambda_{u}^{j}-\lambda_{d}^{j}
    \right]
    dW^{\mathrm{th}}_{n-j-1}\;,
    \\
    \mathcal{N}_{D}^{p}(t,t_{0})
    &=
    \frac{1}{2\Omega}\sqrt{2\gamma n_{\mathrm{th}}}
    \lim_{n\rightarrow\infty}
    \sum_{j=0}^{n-1}
    \left[
        \gamma_{u}\lambda_{u}^{j}
        +
        \gamma_{d}\lambda_{d}^{j}
    \right]
    dW^{\mathrm{th}}_{n-j-1}\;.
\end{align}

The covariance matrix then follows directly from the phase-space solution. Assuming that the stochastic noise has zero mean and is uncorrelated with the initial state, we obtain
\begin{align}
    V_{D}(t)
    &=
    \left\langle
        \mathbf{u}_{D}(t)
        \mathbf{u}_{D}^{T}(t)
    \right\rangle=
    e^{-\gamma t}
    \mathcal{O}_{D}(\Omega t)
    V_{D}(0)
    \mathcal{O}_{D}^{T}(\Omega t)
    +
    \mathcal{Q}_{D}(t,t_{0}),
\end{align}
where $V_{D}(0)$ is the covariance matrix at the beginning of the dark-fringe excursion and
\begin{align}
    \mathcal{Q}_{D}(t,t_{0})
    =
    \left\langle
        \mathbf{N}_{D}(t,t_{0})
        \mathbf{N}_{D}^{T}(t,t_{0})
    \right\rangle=\begin{bmatrix}
        \Big\langle\Big(\mathcal{N}_{D}^{q}(t,t_{0})\Big)^{2}\Big\rangle & \Big\langle\mathcal{N}_{D}^{q}(t,t_{0})\mathcal{N}_{D}^{p}(t,t_{0})\Big\rangle\\
        \Big\langle\mathcal{N}_{D}^{q}(t,t_{0})\mathcal{N}_{D}^{p}(t,t_{0})\Big\rangle & \Big\langle\Big(\mathcal{N}_{D}^{p}(t,t_{0})\Big)^{2}\Big\rangle
    \end{bmatrix}\;.\label{eq_dark_diffusion}
\end{align}
is the covariance accumulated from thermal diffusion during the evolution. The first term therefore describes the deterministic hyperbolic transformation of the initial motional state, including mechanical damping, whereas $\mathcal{Q}_{D}$ accounts for the stochastic broadening generated by thermal noise. This decomposition provides the basis for analyzing the evolution of the principal variances and the resulting dark-fringe thermomechanical squeezing.contribution to the covariance matrix.

The diffusion covariance matrix is obtained by evaluating the second moments of the stochastic contributions in $\mathbf{N}_{D}(t,t_{0})$. Using the zero-mean, uncorrelated Wiener increments and the It\^o relation $\left\langle dW_{i}^{\mathrm{th}}dW_{j}^{\mathrm{th}}\right\rangle=\delta_{ij}\,dt$, each matrix element follows from the corresponding products of the noise kernels in $\mathcal{N}_{D}^{q}$ and $\mathcal{N}_{D}^{p}$. This yields the diffusion contribution to the variances of the position and momentum quadratures and their cross-correlation, which can then be evaluated explicitly from the sums appearing above. Their final expressions take the following form
\begin{align}
    \Big\langle\Big(\mathcal{N}_{D}^{q}(t,t_{0})\Big)^{2}\Big\rangle &=n_{\mathrm{th}}e^{-\gamma t}\Bigg[\Bigg(\frac{\gamma}{2\Omega}\Bigg)\sinh\Big(2\Omega t\Big)+\Bigg(\frac{\gamma}{2\Omega}\Bigg)^{2}\cosh\Big(2\Omega t\Big)\Bigg]-n_{\mathrm{th}}\Bigg[1-e^{-\gamma t}\Bigg]\;,\label{eq_dark_position_variance}\\
    \Big\langle\Big(\mathcal{N}_{D}^{p}(t,t_{0})\Big)^{2}\Big\rangle &=n_{\mathrm{th}}e^{-\gamma t}\Bigg[\Bigg(\frac{\gamma}{2\Omega}\Bigg)\sinh\Big(2\Omega t\Big)-\Bigg(\frac{\gamma}{2\Omega}\Bigg)^{2}\cosh\Big(2\Omega t\Big)\Bigg]+n_{\mathrm{th}}\Bigg[1-e^{-\gamma t}\Bigg]\;,\label{eq_dark_momentum_variance}\\
     \Big\langle\mathcal{N}_{D}^{q}(t,t_{0})\mathcal{N}_{D}^{p}(t,t_{0})\Big\rangle &=\Big(n_{\mathrm{th}}e^{-\gamma t}\Big)\Bigg(\frac{\gamma }{2\Omega}\Bigg)\Bigg[\cosh\Big(2\Omega t\Big)-1\Bigg]\;.\label{eq_dark_cross_corr}
\end{align}
For an initial thermal state, the covariance matrix is isotropic and given by
$V_{D}(0)=n_{\mathrm{th}}\mathcal{I}$.
The full covariance matrix in the dark fringe is then obtained by combining the deterministic hyperbolic evolution with the diffusion contribution through the covariance-matrix relation above. Substituting the initial thermal covariance and the corresponding deterministic and diffusion terms yields the individual covariance elements for an underdamped oscillator
\begin{align}
   V_{D}^{qq}(t)&=n_{\mathrm{th}}e^{-\gamma t}\cosh\Big(2\Omega t\Big)+n_{\mathrm{th}}e^{-\gamma t}\Big(\frac{\gamma}{2\Omega}\Big)\sinh\Big(2\Omega t\Big)-n_{\mathrm{th}}\Big[1-e^{-\gamma t}\Big]\;,\label{eq_dark_Vqq}\\
      V_{D}^{pp}(t)&=n_{\mathrm{th}}e^{-\gamma t}\cosh\Big(2\Omega t\Big)+n_{\mathrm{th}}e^{-\gamma t}\Big(\frac{\gamma}{2\Omega}\Big)\sinh\Big(2\Omega t\Big)+n_{\mathrm{th}}\Big[1-e^{-\gamma t}\Big]\;,\label{eq_dark_Vpp}\\
      V_{D}^{qp}(t)&=n_{\mathrm{th}}e^{-\gamma t}\sinh\Big(2\Omega t\Big)+n_{\mathrm{th}}e^{-\gamma t}\Big(\frac{\gamma }{2\Omega}\Big)\Bigg[\cosh\Big(2\Omega t\Big)-1\Bigg]\;. \label{eq_dark_Vqp}
\end{align}

The limiting behavior of these covariance elements provides useful consistency checks on the dark-fringe solution. At the initial time $t=0$, Eqs.~\eqref{eq_dark_Vqq}--\eqref{eq_dark_Vqp} give
\begin{align}
    V_{D}^{qq}(0)
    &=
    V_{D}^{pp}(0)
    =
    n_{\mathrm{th}},
    \qquad
    V_{D}^{qp}(0)=0,
\end{align}
and hence recover the assumed thermal covariance
\begin{align}
    V_{D}(0)=n_{\mathrm{th}}\mathcal{I}.
\end{align}

For the long-time behavior, $\Omega t\gg1$, 
in the dark-trap regime,  the unstable dynamics overcome the overall damping. The leading asymptotic behavior is therefore
\begin{align}
    V_{D}^{qq}(t)
    &\simeq
    \frac{n_{\mathrm{th}}}{2}
    \left(1+\frac{\gamma}{2\Omega}\right)
    e^{(2\Omega-\gamma)t}\simeq\frac{n_{\mathrm{th}}}{2}
    \left(1+\frac{\gamma}{2\Omega}\right)
    e^{2\Omega t}\;,
    \\
    V_{D}^{pp}(t)
    &\simeq
    \frac{n_{\mathrm{th}}}{2}
    \left(1+\frac{\gamma}{2\Omega}\right)
    e^{(2\Omega-\gamma)t}\simeq \frac{n_{\mathrm{th}}}{2}
    \left(1+\frac{\gamma}{2\Omega}\right)
    e^{2\Omega t},
\end{align}
while
\begin{align}
    V_{D}^{qp}(t)
    &\simeq
    \frac{n_{\mathrm{th}}}{2}
    \left(1+\frac{\gamma}{2\Omega}\right)
    e^{(2\Omega-\gamma)t}\simeq
    \frac{n_{\mathrm{th}}}{2}
    \left(1+\frac{\gamma}{2\Omega}\right)
    e^{2\Omega t}\;.
\end{align}

Thus, all three covariance elements acquire the same dominant exponential scaling, $e^{2\Omega t}$, reflecting the unstable hyperbolic evolution generated by the inverted dark potential. This exponential growth characterizes the thermomechanical evolution of the covariance matrix during the dark-fringe excursion.

We analyze the generation and evolution of phase-space squeezing produced by rapidly transferring the particle between the stable bright fringe and the inverted dark fringe of the optical potential. We first consider an instantaneous switching to dark potential, which provides the basic mechanism and analytical benchmark for squeezing. We then extend the treatment to a finite-duration dark-fringe pulse, where the squeezed state evolves under the inverted potential before returning to the stable configuration. Finally, we consider sequences of dark-fringe pulses and show how repeated switching can control the transient squeezing dynamics.

\subsection{Instantaneous switching to inverted dark potential}
We first consider an instantaneous transfer of the particle from the stable bright trap to the inverted dark potential at time $t=t_{1}$. This is achieved through adaptive phase control by setting the relative phase $\phi = \pi/2$ between counter-propagating fields. The oscillator evolves in the bright trap for $0\leq t<t_1$, while the restoring force is reversed for $t\geq t_1$. Since the switching is instantaneous, the phase-space coordinates $q$ and $p$ do not change discontinuously at the switching time. Consequently, the covariance matrix is continuous across the switching event,
$V(t_1^+)=V(t_1^-)$.

Before the switch, the covariance evolves under stable-trap dynamics. With $\tau_1 = t - t_0$, the propagator for the bright interval is
\begin{equation}
V(t)\big|_{0\leq t<t_1}
=
e^{-\gamma\tau_1}
\mathcal{R}(\Omega\tau_1)
V(0)
\mathcal{R}^{T}(\Omega\tau_1)
+
n_{\mathrm{th}}
\left(1-e^{-\gamma\tau_1}\right)I,
\end{equation}
where $\mathcal{R}$ describes the coherent harmonic rotation of duration $\tau_1$ at frequency $\Omega$.

For an initially thermal state, $V(0)=n_{\mathrm{th}}I$. Since $\mathcal{R}(\Omega\tau_1)
\mathcal{R}^{T}(\Omega\tau_1)
=I$, the thermal covariance is invariant under this rotation. Hence, 
\begin{align}
    V(t)\big|_{0\leq t<t_1} = n_{\mathrm{th}}I\;.
\end{align}
Thus, the stable trap preserves the circular thermal phase-space distribution: harmonic rotation alone does not generate squeezing, while thermal diffusion maintains the equilibrium covariance.

Since the covariance matrix remains continuous across switching from bright to dark fringe, we have
\begin{align}
    V(t_{1}^{-})=V(t_{1}^{+})=n_{\mathrm{th}}I\;.
\end{align}

For the subsequent dark-trap evolution over $t\geq t_{1}$, we define $\tau_2=t-t_1$. The covariance matrix is then
\begin{equation}
V(t)\big|_{t\geq t_1}
=
e^{-\gamma\tau_2}
\mathcal{O}_{D}(\Omega\tau_2)
V(t_1^+)
\mathcal{O}_{D}^{T}(\Omega\tau_2)
+
\mathcal{Q}_{D}(\tau_2)\;,
\end{equation}
where $\mathcal{O}_{D}(\Omega\tau_{2})$ is the deterministic hyperbolic propagator of the inverted trap [Eq.~\eqref{eq_dark_deterministic_sol}] and $\mathcal{Q}_{D}(\tau_{2})$ represents the thermal covariance accumulated during the dark-trap evolution  [Eq.~\eqref{eq_dark_diffusion}]. In the underdamped regime, the covariance matrix elements for $t \geq t_1$ are given by Eqs.~\eqref{eq_dark_Vqq}-\eqref{eq_dark_Vqp}.

The essential difference between the bright and dark traps is therefore the change from trigonometric to hyperbolic phase-space evolution. In the bright trap, the distribution undergoes ordinary rotation, whereas in the dark trap one phase-space direction is exponentially amplified while the conjugate direction is compressed. An initially circular thermal distribution consequently develops an elliptical shape.

\subsection{Finite dark fringe pulse}
We now derive the covariance dynamics for a single finite-duration dark-trapping pulse.  The protocol consists of three stages: bright trap ($0 \leq t < t_1$), dark trap ($t_1 \leq t < t_2$), and bright trap ($t \geq t_2$). The dark-pulse duration is
\begin{equation}
    \mathcal{T}_{21} = t_2 - t_1\;.
\end{equation}

We assume instantaneous switching at both $t_1$ and $t_2$, so the covariance matrix is continuous across each switching event:
\begin{equation}
    V(t_1^+) = V(t_1^-), \qquad V(t_2^+) = V(t_2^-)\;.
\end{equation}

\subsubsection{Bright-trap evolution before the pulse}

Before the first switching, the system evolves in the bright fringe. For an initially thermal state, $V(0) = n_{\mathrm{th}} I$, the covariance remains
\begin{equation}
    V(t) = n_{\mathrm{th}} I, \qquad 0 \leq t < t_1\;.\label{eq_dark_pulse_step_1}
\end{equation}
Thus, immediately before the first switching,
\begin{equation}
    V(t_1^-) = n_{\mathrm{th}} I\;,
\end{equation}
and hence $V(t_1^+) = n_{\mathrm{th}} I$.

\subsubsection{Dark-trap dynamics}

During the dark interval $t_1 \leq t < t_2$, the inverted potential amplifies displacements. With $\tau_2 = t - t_1$, the covariance evolves as
\begin{equation}
    V(t)
    =
    e^{-\gamma\tau_2}
    \mathcal{O}_{D}(\Omega\tau_2)
    V(t_1^+)
    \mathcal{O}_{D}^{T}(\Omega\tau_2)
    +
    \mathcal{Q}_{D}(\tau_2)\;,
\end{equation}
where $\mathcal{O}_{D}$ is the hyperbolic propagator of the inverted trap and $\mathcal{Q}_{D}$ accounts for thermal noise during the dark interval.

At the end of the dark pulse, we define
\begin{equation}
    V_D(\mathcal{T}_{21})
    \equiv V(t_2^-)
    =
    \begin{bmatrix}
        V_D^{qq}(\mathcal{T}_{21}) & V_D^{qp}(\mathcal{T}_{21})\\
        V_D^{pq}(\mathcal{T}_{21}) & V_D^{pp}(\mathcal{T}_{21})
    \end{bmatrix}\;.
\end{equation}

The covariance elements are
\begin{align}
    V_{D}^{qq}(\mathcal{T}_{21})
    &= n_{\mathrm{th}} e^{-\gamma \mathcal{T}_{21}} \cosh(2\Omega \mathcal{T}_{21})
    + n_{\mathrm{th}} e^{-\gamma \mathcal{T}_{21}}
        \frac{\gamma}{2\Omega} \sinh(2\Omega \mathcal{T}_{21})
    - n_{\mathrm{th}} \left(1 - e^{-\gamma \mathcal{T}_{21}}\right)\;,
\end{align}
\begin{align}
    V_{D}^{pp}(\mathcal{T}_{21})
    &= n_{\mathrm{th}} e^{-\gamma \mathcal{T}_{21}} \cosh(2\Omega \mathcal{T}_{21})
    + n_{\mathrm{th}} e^{-\gamma \mathcal{T}_{21}}
        \frac{\gamma}{2\Omega} \sinh(2\Omega \mathcal{T}_{21})
    + n_{\mathrm{th}} \left(1 - e^{-\gamma \mathcal{T}_{21}}\right)\;,
\end{align}
\begin{align}
    V_{D}^{qp}(\mathcal{T}_{21})
    &= n_{\mathrm{th}} e^{-\gamma \mathcal{T}_{21}} \sinh(2\Omega \mathcal{T}_{21})
    + n_{\mathrm{th}} e^{-\gamma \mathcal{T}_{21}}
    \Bigg(\frac{\gamma}{2\Omega}\Bigg)
    \Bigg[
        \cosh(2\Omega \mathcal{T}_{21}) - 1
    \Bigg]\;.
\end{align}

\subsubsection{Bright-trap evolution after the pulse}

The second switching restores the harmonic trap in the bright fringe, with $V(t_2^+) = V_D(\mathcal{T}_{21})$. For $t \geq t_2$, with $\tau_3 = t - t_2$, the covariance evolves as
\begin{equation}
    V(t)
    =
    e^{-\gamma\tau_3}
    \mathcal{R}(\Omega\tau_3)
    V_D(\mathcal{T}_{21})
    \mathcal{R}^{T}(\Omega\tau_3)
    +
    n_{\mathrm{th}}
    \left(1 - e^{-\gamma\tau_3}\right) I\;,
\end{equation}
where $\mathcal{R}(\Omega\tau_3)$ is the stable-trap rotation matrix.

The resulting covariance elements are
\begin{align}
    V_{qq}(t)
    &=
    e^{-\gamma\tau_3}
    \Big[
        V_D^{qq}(\mathcal{T}_{21}) \cos^2(\Omega\tau_3)
        + V_D^{pp}(\mathcal{T}_{21}) \sin^2(\Omega\tau_3)
        + V_D^{qp}(\mathcal{T}_{21}) \sin(2\Omega\tau_3)
    \Big] \nonumber\\
&\quad
    + n_{\mathrm{th}} \left(1 - e^{-\gamma\tau_3}\right)\;,
\end{align}
\begin{align}
    V_{pp}(t)
    &=
    e^{-\gamma\tau_3}
    \Big[
        V_D^{qq}(\mathcal{T}_{21}) \sin^2(\Omega\tau_3)
        + V_D^{pp}(\mathcal{T}_{21}) \cos^2(\Omega\tau_3)
        - V_D^{qp}(\mathcal{T}_{21}) \sin(2\Omega\tau_3)
    \Big] \nonumber\\
&\quad
    + n_{\mathrm{th}} \left(1 - e^{-\gamma\tau_3}\right)\;,
\end{align}
\begin{align}
    V_{qp}(t)
    &=
    e^{-\gamma\tau_3}
    \Bigg[
        \frac{V_D^{pp}(\mathcal{T}_{21}) - V_D^{qq}(\mathcal{T}_{21})}{2}
        \sin(2\Omega\tau_3)
        + V_D^{qp}(\mathcal{T}_{21}) \cos(2\Omega\tau_3)
    \Bigg]\;.
\end{align}
The above equations show that the dark-trapping pulse determines the covariance matrix
$V_D(\mathcal{T}_{21})$, while the subsequent bright-trap evolution rotates and damps this covariance ellipse. In particular, the quantities
$V_D^{qq}(\mathcal{T}_{21})$, $V_D^{pp}(\mathcal{T}_{21})$, and  $V_D^{qp}(\mathcal{T}_{21})$
contain the complete information about the squeezing generated during the dark pulse.

\subsubsection{Squeezing and relaxation}
The principal variances are obtained from the eigenvalues $\Lambda_{\mp}$ of the covariance matrix, with the squeezing level defined in Eq.~\eqref{eq_squeezing_dB}. At $t = t_2^+$, the covariance is $V_D(\mathcal{T}_{21})$. For $t > t_2$, the stable dynamics rotate the squeezed ellipse as its anisotropy decays to equilibrium. In the long-time limit,
\begin{equation}
    V(t) \xrightarrow[t \rightarrow \infty]{} n_{\mathrm{th}} I,
    \qquad
    \mathcal{S}_{\mathrm{dB}} \xrightarrow[t \rightarrow \infty]{} 0\;.
\end{equation}

\subsubsection{Numerical simulation}
To illustrate the squeezing effect via the hyperbolic evolution of the particle in the dark fringe, the numerical solution~\cite{PhysRevA.45.600} of Eqs.~\eqref{eq_dark_diff_q} and \eqref{eq_dark_diff_p} is obtained using a time step $dt = 4 \times 10^{-5}/\gamma$, selected for numerical stability. The Wiener increments are implemented as $dW_{\rm th}(t) = \sqrt{dt} \, \mathcal{N}(0, 1)$, with $\mathcal{N}(0, 1)$ being the standard normal distribution.

\begin{figure}[ht!]
    \centering
     \includegraphics[width=\linewidth]{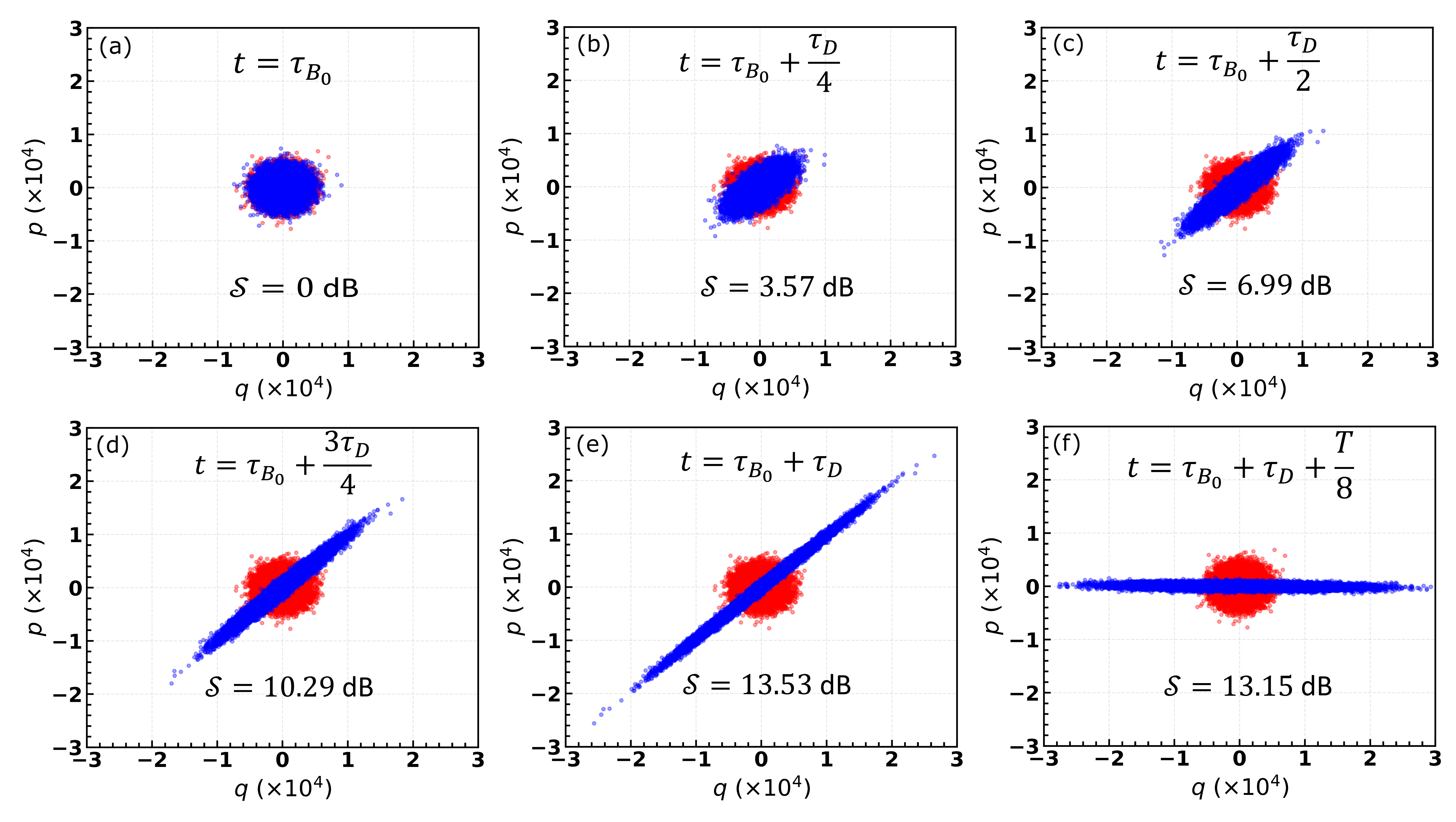}
   \caption{
    \textbf{Hyperbolic phase-space compression during a dark-fringe excursion.}
    The inverted optical potential at the dark fringe produces exponential squeezing
    that grows monotonically over the pulse duration $\tau_D = T/4$.
    (a) Isotropic thermal state at $t = \tau_{B_0}$.
    (b) 25\% traversal ($t = \tau_{B_0} + \tau_D/4$), $\mathcal{S} = 3.57$~dB.
    (c) 50\% ($t = \tau_{B_0} + \tau_D/2$), $\mathcal{S} = 6.99$~dB.
    (d) 75\% ($t = \tau_{B_0} + 3\tau_D/4$), $\mathcal{S} = 10.29$~dB.
    (e) Pulse end ($t = \tau_{B_0} + \tau_D$), maximum $\mathcal{S} = 13.53$~dB.
    (f) $T/8$ post-pulse delay, $\mathcal{S} = 13.15$~dB, reflecting the onset of
    dissipative decay. The snapshots are generated from 20,000 trajectories, with initial conditions randomly sampled from a Boltzmann distribution for a cold thermal state with occupation number  $n_{\mathrm{th}}$.
}\label{Fig_SM_dark_pulse_numerical}
\end{figure}

The dark-fringe protocol generates squeezing through a fundamentally different mechanism than the frequency-quench protocol. Rather than an instantaneous symplectic transformation, the inverted optical potential at the dark fringe
drives continuous hyperbolic evolution in which one quadrature is exponentially compressed while the other is exponentially stretched. Fig.~\ref{Fig_SM_dark_pulse_numerical} tracks this process in real time. At the start of the pulse (a), the distribution is isotropic. As the particle traverses the dark fringe, the squeezing builds continuously: $\mathcal{S} = 3.57$~dB at 25\% (b), $6.99$~dB at 50\% (c), $10.29$~dB at 75\% (d), and $13.53$~dB at the pulse end (e). The near-linear growth in dB reflects the exponential compression in phase space. Following the return to the bright fringe, the phase-space distribution rotates under the influence of damping and thermal diffusion. One such snapshot, taken at $T/8$ after the pulse, shows a squeezing of $13.15$~dB (f). Throughout this
evolution, the numerically computed squeezing values agree excellently with the
analytical predictions for hyperbolic dynamics.

\subsection{Sequence of dark fringe pulses}
We now generalize the single dark-pulse protocol to a sequence of $N$ finite dark-trap pulses. The protocol consists of alternating dark and bright trapping intervals,
\[
B \rightarrow D_1 \rightarrow B_1 \rightarrow D_2 \rightarrow \cdots \rightarrow B_{N-1} \rightarrow D_N \rightarrow B,
\]
where $D_j$ denotes the $j$-th dark pulse and $B_j$ denotes the bright interval between the $j$-th and $(j+1)$-th dark pulses. The duration of the $j$-th dark pulse is $\tau_{D,j}$, while the subsequent bright interval has duration $\tau_{B,j}$.

Let $V_j^-$ and $V_j^+$ denote the covariance matrices immediately before and after the $j$-th dark pulse, respectively. The covariance evolves according to the general form
\begin{equation}
    V(t)
    =
    e^{-\gamma \tau}
    \mathcal{O}(t,t_0)
    V(t_0)
    \mathcal{O}^{T}(t,t_0)
    +
    \mathcal{Q}(t,t_0)\;,
\end{equation}
where $\tau = t - t_0$, $\mathcal{O}$ describes the deterministic evolution, and $\mathcal{Q}$ accounts for the covariance generated by stochastic noise.

\subsubsection{Dark-trap evolution}

During a dark interval, the deterministic propagator is the hyperbolic matrix
\begin{equation}
    \mathcal{O}_{D}(\Omega\tau_D)
    =
    \begin{bmatrix}
        \cosh(\Omega\tau_D) & \sinh(\Omega\tau_D)\\
        \sinh(\Omega\tau_D) & \cosh(\Omega\tau_D)
    \end{bmatrix}\;.
\end{equation}

The covariance immediately after the $j$-th dark pulse is therefore
\begin{equation}
    V_j^+
    =
    e^{-\gamma\tau_{D,j}}
    \mathcal{O}_{D}(\Omega\tau_{D,j})
    V_j^-
    \mathcal{O}_{D}^{T}(\Omega\tau_{D,j})
    +
    \mathcal{Q}_{D}(\tau_{D,j})\;,
\end{equation}
where $\mathcal{Q}_{D}(\tau_{D,j})$ is the thermal covariance accumulated during the dark interval.

\subsubsection{Bright-trap evolution}

Following the $j$-th dark pulse, the system is switched to the stable bright trap. The corresponding deterministic propagator is the rotation matrix
\begin{equation}
    \mathcal{O}_{B}(\Omega\tau_B)
    =
    \begin{bmatrix}
        \cos(\Omega\tau_B) & \sin(\Omega\tau_B)\\
        -\sin(\Omega\tau_B) & \cos(\Omega\tau_B)
    \end{bmatrix}\;.
\end{equation}

The covariance after the subsequent bright interval is
\begin{equation}
    V_{j+1}^-
    =
    e^{-\gamma\tau_{B,j}}
    \mathcal{O}_{B}(\Omega\tau_{B,j})
    V_j^+
    \mathcal{O}_{B}^{T}(\Omega\tau_{B,j})
    +
    \mathcal{Q}_{B}(\tau_{B,j})\;,
\end{equation}
where the bright-trap noise covariance is isotropic:
\begin{equation}
    \mathcal{Q}_{B}(\tau_{B,j})
    =
    n_{\mathrm{th}}
    \left(1 - e^{-\gamma\tau_{B,j}}\right) I\;.
\end{equation}

\subsubsection{Cycle recurrence}

For a sequence of identical dark pulses of duration $\tau_D$ separated by identical bright intervals of duration $\tau_B$, we define one complete dark-bright cycle by the deterministic map
\begin{equation}
    \mathcal{M}
    =
    \mathcal{O}_{B}(\Omega\tau_B)
    \mathcal{O}_{D}(\Omega\tau_D)\;,
\end{equation}
where the ordering reflects that the dark propagator acts first, followed by the bright propagator.

Combining the dark and bright evolutions and including the accumulated noise, the covariance after one complete cycle is
\begin{equation}
    V_{j+1}
    =
    A V_j A^T
    +
    \mathcal{Q}_{\mathrm{cyc}}\;,\label{eq_cov_after_one_cycle}
\end{equation}
where we define
\begin{equation}
    A
    =
    e^{-\gamma(\tau_D+\tau_B)/2}
    \mathcal{M}\;,
\end{equation}
and the total noise generated during one cycle is
\begin{equation}
    \mathcal{Q}_{\mathrm{cyc}}
    =
    e^{-\gamma\tau_B}
    \mathcal{O}_{B}(\Omega\tau_B)
    \mathcal{Q}_{D}(\tau_D)
    \mathcal{O}_{B}^{T}(\Omega\tau_B)
    +
    \mathcal{Q}_{B}(\tau_B)\;.
\end{equation}

\subsubsection{Closed-form solution for $N$ cycles}

The recurrence in Eq.~\eqref{eq_cov_after_one_cycle} can be iterated to obtain a closed-form expression for the covariance after $N$ cycles:
\begin{equation}
    V_N
    =
    A^N V_0 (A^T)^N
    +
    \sum_{k=0}^{N-1}
    A^k
    \mathcal{Q}_{\mathrm{cyc}}
    (A^T)^k\;.
\end{equation}

Substituting $A = e^{-\gamma(\tau_D+\tau_B)/2}\mathcal{M}$, this becomes
\begin{equation}
    V_N
    =
    e^{-N\gamma(\tau_D+\tau_B)}
    \mathcal{M}^N
    V_0
    (\mathcal{M}^T)^N
    +
    \sum_{k=0}^{N-1}
    e^{-k\gamma(\tau_D+\tau_B)}
    \mathcal{M}^k
    \mathcal{Q}_{\mathrm{cyc}}
    (\mathcal{M}^T)^k\;.
\end{equation}

The physical interpretation is clear: the first term represents the initial covariance after $N$ deterministic cycles, while the sum represents the accumulated stochastic noise, with earlier noise contributions propagated through more subsequent cycles. Thus, the effect of repeated pulses cannot be obtained simply by multiplying single-pulse squeezing by $N$.

\subsubsection{Bright-interval phase matching}

The bright-trap duration $\tau_B$ is a crucial control parameter. During the bright interval, the covariance ellipse rotates by
\begin{equation}
    \theta_B = \Omega\tau_B\;.
\end{equation}

For $\tau_B = T = 2\pi/\Omega$, the rotation angle is $2\pi$, returning the ellipse to its original orientation. For $\tau_B = T/2$, the rotation angle is $\pi$, which also preserves the ellipse orientation (since an ellipse is invariant under a $\pi$ rotation). For $\tau_B = T/4$, the rotation angle is $\pi/2$, which interchanges the squeezed and antisqueezed directions, generally reducing the effectiveness of subsequent dark pulses.

% However, the optimal $\tau_B$ cannot be determined from the rotation angle alone. The full covariance recurrence must be considered:
% \begin{equation}
%     V_{j+1}
%     =
%     e^{-\gamma(\tau_D+\tau_B)}
%     \mathcal{M}
%     V_j
%     \mathcal{M}^{T}
%     +
%     \mathcal{Q}_{\mathrm{cyc}}\;,
% \end{equation}
% where both $\mathcal{M}$ and $\mathcal{Q}_{\mathrm{cyc}}$ depend on $\tau_B$. Changing $\tau_B$ modifies not only the ellipse orientation but also how fluctuations are rotated, damped, and accumulated.

% The optimal bright interval is therefore defined by minimizing the smallest eigenvalue of the final covariance matrix:
% \begin{equation}
%     \tau_{B,\mathrm{opt}}
%     =
%     \underset{\tau_B}{\operatorname{arg\,min}}
%     \left\{
%         \lambda_{\min}\left[V_N(\tau_B)\right]
%     \right\}\;.
% \end{equation}

\subsubsection{Squeezing generated by the pulse sequence}

The squeezing is determined by the smallest eigenvalue of the covariance matrix and can be obtained using Eq.~\eqref{eq_eigenvalues}. The squeezing in decibels is then governed by Eq.~\eqref{eq_squeezing_dB}. In the ideal limit of negligible damping and noise, each dark pulse reduces the squeezed variance by $e^{-2\Omega\tau_D}$. With appropriate timing between pulses, the squeezed variance after $N$ pulses scales as
\begin{equation}
    \Lambda_{\min}^{(N)}
    \simeq
    n_{\mathrm{th}}
    e^{-2N\Omega\tau_D}\;,
\end{equation}
corresponding to
\begin{equation}
    \mathcal{S}_N
    \simeq
    -\frac{20\Omega\tau_D}{\ln 10}\,N\;.
\end{equation}

In the realistic dissipative system, this ideal scaling is modified by damping and stochastic diffusion. The deviation from exact $N$-fold squeezing is determined by the competition between the exponential amplification during each dark pulse and the covariance added by $\mathcal{Q}_{D}(\tau_D)$ and $\mathcal{Q}_{B}(\tau_B)$ throughout the sequence.

\section{Nonlinear optomechanics and preparation of non-Gaussian state}
\label{appendixE}

To preserve the full optical nonlinearity, we avoid expanding the standing-wave intensity about the equilibrium position, as was done in the earlier force equation (Eq.~\eqref{eq_optical_force}). Instead, we keep all orders of the expansion and write the most general form of the axial gradient force as 
\begin{align}
    F( z)=
    \frac{\alpha'}{4}
    \frac{\partial |E(z,t)|^2}{\partial z}
    =
    -\alpha'|\mathcal{E}_0|^2k_l
    \sin(2k_lz).
    \label{eq:exact_optical_force}
\end{align}
This expression is the exact gradient force within the standing-wave model and therefore retains the complete spatial nonlinearity contained in the sinusoidal term. 

We characterize the optical nonlinearity with the dimensionless Lamb--Dicke-like parameter
\begin{align}
    \eta
    \equiv
    k_l z_{\rm zpm}
    =
    k_l\sqrt{\frac{\hbar}{m\Omega}},
    \label{eq:eta_definition}
\end{align}
which compares the zero-point displacement to the standing-wave period. The force prefactor can then be written in terms of frequency as
\begin{align}
    \frac{\alpha'|\mathcal{E}_0|^2k_l}{p_{\rm zpm}}
    =
    \frac{\Omega_0}{2\eta}.
    \label{eq:force_prefactor}
\end{align}
Using this, the dimensionless equations of motion therefore take the compact form
\begin{align}
    d q
    &=
    \Omega_0 p\,dt,
    \label{eq_nonlinear_q}\\
    d p
    &=
    \left[
        -\frac{\Omega_0}{2\eta}
        \sin(2\eta q)
        -\gamma p
    \right]dt
    +
    dW^{\rm th}(t)\;, 
    \label{eq_nonlinear_p}
\end{align}
where $dW^{\rm th}(t)$ thermal Wiener increments. In the small-displacement limit, $\eta q\ll1$, we retain the usual harmonic Langevin dynamics.  The sine dependence in
Eq.~\eqref{eq_nonlinear_p}, retains all higher-order nonlinear corrections of the optical force.

\subsubsection{Expansion about bright and dark fringes}
\label{sec:bright_dark}

The distinction between bright- and dark-fringe operation is most transparent
when the local optical phase is used as the expansion variable.  Defining
\begin{align}
    2k_l z + 2\phi(t) = \tilde{z}
\end{align}
for a bright fringe, the force
$F=-K\sin \tilde{z}$, with $\kappa_{\rm trap}\equiv\alpha'|\mathcal{E}_0|^2k_l$,
has the expansion
\begin{align}
    F_{\rm bright}
    &=
    -\kappa_{\rm trap}\sin \tilde{z}=
    -\kappa_{\rm trap}\tilde{z}+\frac{\kappa_{\rm trap}}{6}\tilde{z}^3+\mathcal O(\tilde{z}^5).
    \label{eq:bright_force_expansion}
\end{align}
Thus, the leading nonlinear correction is cubic and has the opposite sign to the linear restoring term.

At a dark fringe, the local phase is shifted by $\pi$, so that
\begin{align}
    2k_l z + 2\phi(t)=\pi+\tilde{z} .
\end{align}
Using $\sin(\pi+\tilde{z})=-\sin \tilde{z}$, one obtains
\begin{align}
    F_{\rm dark}
    &=
    \kappa_{\rm trap}\sin \tilde{z}=
    \kappa_{\rm trap}\tilde{z}-\frac{\kappa_{\rm trap}}{6}\tilde{z}^3+\mathcal O(\tilde{z}^5).
    \label{eq:dark_force_expansion}
\end{align}
Hence both the linear and cubic terms change sign between the two operating points.  For $\alpha'>0$, this reflects the reversal of the local curvature of the optical potential: the bright fringe is locally stable, whereas the
dark fringe is locally unstable.  Importantly, these polynomial expansions are only local descriptions; the full sinusoidal force in Eq.~\eqref{eq:exact_optical_force} remains valid beyond the small-displacement regime.

\section{Recoil heating}
\label{appendixF}
\subsection{For polarization-adaptive squeezing}
A key advantage of the polarization-adaptive squeezing protocol is the ability to control the trap stiffness without changing the total optical trapping power. At the same time, this control introduces an important consideration for operation at lower pressures: the expected increase in backaction heating as the pressure is reduced toward ultra-high vacuum~\cite{Recoil_Jain}. %For two counterpropagating beams of equal power $P$ and relative polarization angle $\theta_p$, the standing-wave trap depth scales as $U\propto P|\cos{\theta_p}|$, such that, at fixed $P$, the axial trap frequency scales as $\Omega_z\propto\sqrt{|\cos{\theta_p}|}$ (see SM). In contrast, the recoil heating rate~\cite{Recoil_Jain} $\Gamma_{\rm rec}$, expressed in phonons per unit time, scales as $\Gamma_{\rm rec}/\Gamma_{\rm rec, 0}\propto P/\Omega_z \sim 0.5\times(1+|\cos{\theta_p}|)/\sqrt{|\cos{\theta_p}|}$ (see SM). For $\theta_p=88.3^\circ$, this yields $\Omega_z/\Omega_{z,0}=0.17$ and $\Gamma_{\rm rec}/\Gamma_{\rm rec, 0}=4$, where the subscript $0$ denotes parallel polarization.
The recoil heating is determined by photon scattering from the local optical field~\cite{Recoil_Jain}. Assuming plane-wave like core modes, the cycle-averaged intensity is related to the complex field amplitude $E(z)$ as given in Eq.~\eqref{eq_electric_field}:
\begin{align}
I(z)
=
\frac{1}{2}\epsilon_0 c|E(z)|^2.
\label{eq:field_intensity}
\end{align}
Near the equilibrium bright fringe position $z=0$, Eq.~\eqref{eq:field_intensity} therefore gives
\begin{align}
I_{\rm b}
&=
\epsilon_0c|\mathcal{E}_0|^2
\left(
1+|\cos\theta_p|
\right).
\label{eq:trap_intensity}
\end{align}
Thus, for fixed power $P$ in each beam,
\begin{align}
I_{\rm b}
\propto
P\left(1+|\cos\theta_p|\right).
\label{eq:intensity_scaling}
\end{align}

The total scattered power is
\begin{align}
P_{\rm sc}
=
\sigma_{\rm sc}I_{\rm b},
\label{eq:scattered_power}
\end{align}
where $\sigma_{\rm sc}$ is the photon-scattering cross section. For a Rayleigh particle the scattering cross-section can be written as~\cite{Recoil_Jain}:
\begin{align}
\sigma_{\rm sc}
=
\frac{k_\ell^4|\alpha|^2}
{6\pi\epsilon_0^2}.
\label{eq:scattering_cross_section}
\end{align}
Since $\sigma_{\rm sc}$ is independent of $\theta_p$ for the particle considered here, Eqs.~\eqref{eq:intensity_scaling} and \eqref{eq:scattered_power} give
\begin{align}
P_{\rm sc}
\propto
P\left(1+|\cos\theta_p|\right).
\label{eq:scattering_scaling}
\end{align}

The corresponding photon-scattering rate is
\begin{align}
R_{\rm sc}
=
\frac{P_{\rm sc}}{\hbar\omega_\ell}.
\end{align}
Photon scattering produces momentum diffusion and hence recoil heating of the mechanical mode. Following Ref.~\cite{Recoil_Jain}, the recoil heating rate of the axial mode, expressed in phonons per unit time, is
\begin{align}
\Gamma_{\rm rec}
=
\frac{1}{5}
\frac{P_{\rm sc}}{mc^2}
\frac{\omega_\ell}{\Omega},
\label{eq:recoil_heating}
\end{align}
where $m$ is the particle mass, $c$ is the speed of light, and $\Omega$ is the axial trap frequency.

Combining Eqs.~\eqref{eq:scattering_scaling}, \eqref{eq:recoil_heating}, and \eqref{eq_frequency}, we find
\begin{align}
\Gamma_{\rm rec}
&\propto
\frac{P(1+|\cos\theta_p|)}
{\Omega}
\nonumber\\
&\propto
\sqrt{P}\,
\frac{1+|\cos\theta_p|}
{\sqrt{|\cos\theta_p|}}.
\end{align}
Therefore, at fixed optical power,
\begin{align}
\Gamma_{\rm rec}
\propto
\frac{1+|\cos\theta_p|}
{\sqrt{|\cos\theta_p|}}\;.
\label{eq:heating_scaling}
\end{align}

Normalizing to the parallel-polarization configuration, $\theta_p=0$, gives
\begin{align}
\frac{\Gamma_{\rm rec}}
{\Gamma_{\rm rec,0}}
=
\frac{1+|\cos\theta_p|}
{2\sqrt{|\cos\theta_p|}}\;.
\label{eq:heating_ratio}
\end{align}
From this, we can find that rotating the relative polarization from parallel to $\theta_p=88.3^\circ$ reduces the axial trap frequency to approximately $11.1\%$ of its value for parallel polarization (denoted with subscript 0), while increasing the recoil heating rate, expressed in axial phonons per unit time, by a factor of approximately $3$. This behavior is distinct from conventional squeezing protocols based on reducing the trapping intensity, for which both the trap stiffness and the recoil energy-diffusion rate decrease linearly with the optical power $P$. The standing-wave geometry, however, offers an additional degree of control that can be used to suppress recoil heating spatially.
\subsection{Recoil suppression near a dark fringe}

An alternative to polarization-based control is to operate the standing-wave trap near a dark fringe, where the local optical intensity is suppressed while the optical potential retains finite curvature. For parallel polarizations, $\theta_p=0$, the standing-wave intensity near the dark fringe is
\begin{align}
|E(z)|^2
=
4|\mathcal{E}_0|^2\sin^2(k_\ell z)
\simeq
4|\mathcal{E}_0|^2k_\ell^2z^2,
\label{eq:dark_field}
\end{align}
where the last equality holds for a feedback-cooled particle in the absence of squeezing as $k_\ell |z|\ll1$.

The recoil heating rate is proportional to the photon-scattering rate and hence to the local intensity. Using
\[
I(z)=\frac{1}{2}\epsilon_0c|E(z)|^2,
\]
the mean intensity sampled near the dark fringe is
\begin{align}
\langle I_d(z)\rangle
\simeq
2\epsilon_0c|\mathcal{E}_0|^2
k_\ell^2\langle z^2\rangle,
\end{align}
where $\langle z^2\rangle$ is the mean-square axial displacement. At a bright fringe,
\begin{align}
I_b=2\epsilon_0c|\mathcal{E}_0|^2.
\end{align}
Since the scattered power is $P_{\rm sc}=\sigma_{\rm sc}I$, the cross section cancels in the ratio, giving
\begin{align}
\frac{\Gamma_d}{\Gamma_b}
=
\frac{\langle I_d(z)\rangle}{I_b}
\simeq
k_\ell^2\langle z^2\rangle
\label{eq:dark_bright_recoil}
\end{align}
where $\Gamma_d$ and $\Gamma_b$ are the recoil heating rates at the dark and bright fringes, respectively.

Thus, in the Lamb--Dicke regime,
\begin{align}
k_\ell^2\langle z^2\rangle\ll1,
\end{align}
recoil heating is strongly suppressed while the trap curvature remains finite. During accelerated squeezing, the increasing motional variance $\langle z^2(t)\rangle$ progressively increases the scattering rate according to
\begin{align}
\Gamma_d(t)
\simeq
\Gamma_b k_\ell^2\langle z^2(t)\rangle.
\end{align}

\begin{figure}[ht!]
    \centering
    \includegraphics[width=0.9\linewidth]{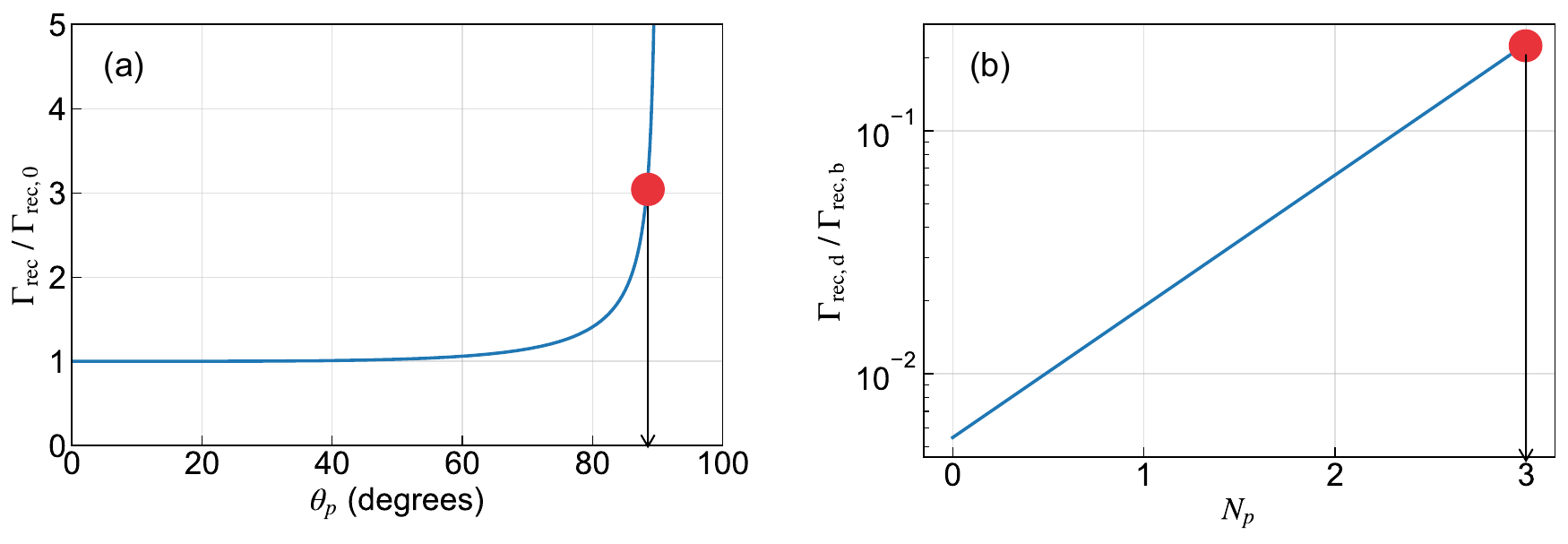}
    \caption{\textbf{Numerically calculated recoil heating;} (a) Analytical recoil heating rate during polarization-adaptive delocalization as a function of the relative polarization angle $\theta_p$ between the trapping beams. The recoil rate at the red circle ($\theta_p=88.3^\circ$) is 3 times higher than that of at $\theta_p=0^\circ$.  (b) Ratio of the analytical recoil-heating rates in the dark and bright fringe configurations, $\Gamma_{\mathrm{rec,d}}/\Gamma_{\mathrm{rec,b}}$, as a function of the number of dark-pulse events $N_p$. Here $N_p=0$ means the particle is just transferred from bright to dark fringe for the first time. The red circle and the black arrow indicate our experiment conditions highlighting the fact that the recoil rate in the dark fringe trap at the maximum delocalization instance of 13 dB is $\sim$0.25 times smaller than that of in the bright fringe trap.}
    \label{fig5}
\end{figure}

During accelerated squeezing, however, the growth of $\langle z^2(t)\rangle$ progressively increases the recoil rate. The resulting increase in recoil heating therefore establishes a trade-off between increasing the degree of squeezing/delocalization and maintaining low recoil heating.  In Fig.~\ref{fig5}(a), we indicate theoretical recoil heating for polarization-adaptive squeezing as a function of $\theta_p$. In Fig.~\ref{fig5}(b) the ratio of the recoil-heating rates in the dark and bright trapping configurations, $\Gamma_{\mathrm{rec,d}}/\Gamma_{\mathrm{rec,b}}$, as a function of the number of dark-trapping events $N$. The red circle and the black arrow indicates our present experimental conditions. 

\twocolumngrid
%\bibliographystyle{apsrev4-2}
%\bibliography{arXiv_ref}

\end{document}